\documentclass[10pt,twocolumn,twoside]{IEEEtran}
\usepackage{amsmath}
\usepackage{amssymb}

\usepackage{graphicx}
\usepackage{epstopdf}
\usepackage{amsmath}
\usepackage{array}
\newcommand{\PreserveBackslash}[1]{\let\temp=\\#1\let\\=\temp}
\newcolumntype{C}[1]{>{\PreserveBackslash\centering}p{#1}}
\newcolumntype{R}[1]{>{\PreserveBackslash\raggedleft}p{#1}}
\newcolumntype{L}[1]{>{\PreserveBackslash\raggedright}p{#1}}
\usepackage[usenames]{color}
\usepackage{colortbl,booktabs}
\usepackage{tabularx}
\usepackage{multicol}
\usepackage{booktabs}
\usepackage[Symbol]{upgreek}
\usepackage{bm}
\usepackage{cite}
\usepackage{balance}
\usepackage{mathrsfs}
\usepackage{url}
\usepackage{threeparttable}
\usepackage{amsmath}
\usepackage{dcolumn}
\usepackage{multirow}
\usepackage{booktabs}
\usepackage{amsfonts}
\usepackage{enumerate}
\usepackage{subfigure}
\usepackage{algorithm}
\usepackage{algorithmic}

\begin{document}
	\title{Weighted Sum Rate Maximization of the mmWave Cell-Free MIMO Downlink Relying on Hybrid Precoding}
	\author{\IEEEauthorblockN{Author list}\vspace{-0mm}}
	\author{Chenghao Feng, Wenqian Shen, Jianping An, and Lajos Hanzo,~\IEEEmembership{Fellow,~IEEE}.
        \thanks{This work was supported in part by the National Natural Science Foundation of China (NSFC) under Grant 61901034, and in part by the Open Research Fund of the Shaanxi Province Key Laboratory of Information Communication Network and Security under Grant ICNS201905.
                L. Hanzo would like to acknowledge the financial support of the Engineering and Physical Sciences Research Council projects EP/N004558/1, EP/P034284/1, EP/P034284/1, EP/P003990/1 (COALESCE), of the Royal Society's Global Challenges Research Fund Grant as well as of the European Research Council's Advanced Fellow Grant QuantCom. (\it{Corresponding author: Wenqian Shen.})}
        \thanks{
		C. Feng, W. Shen, and J. An are with the School of Information and Electronics, Beijing Institute of Technology, Beijing 100081, China (e-mails: cfeng@bit.edu.cn, shenwq@bit.edu.cn, and an@bit.edu.cn).\par
        L. Hanzo is with the Department of Electronics and Computer Science, University of Southampton, Southampton SO17 1BJ, UK (e-mail: lh@ecs.soton.ac.uk).
		}
	}
	\maketitle
	\begin{abstract}
    The cell-free MIMO concept relying on hybrid precoding constitutes an innovative technique capable of dramatically increasing the network capacity of millimeter-wave (mmWave) communication systems.
    It dispenses with the cell boundary of conventional multi-cell MIMO systems, while drastically reducing the power consumption by limiting the number of radio frequency (RF) chains at the access points (APs).
    In this paper, we aim for maximizing the weighted sum rate (WSR) of mmWave cell-free MIMO systems by conceiving a low-complexity hybrid precoding algorithm.
    We formulate the WSR optimization problem subject to the transmit power constraint for each AP and the constant-modulus constraint for the phase shifters of the analog precoders.
    A block coordinate descent (BCD) algorithm is proposed for iteratively solving the problem.
    In each iteration, the classic Lagrangian multiplier method and the penalty dual decomposition (PDD) method are combined for obtaining near-optimal hybrid analog/digital precoding matrices.
    Furthermore, we extend our proposed algorithm for deriving closed-form expressions for the precoders of fully digital cell-free MIMO systems.
    Moreover, we present the convergency analysis and complexity analysis of our proposed method.
    Finally, our simulation results demonstrate the superiority of the algorithms proposed for both fully digital and hybrid precoding matrices.
	\end{abstract}
	
	\IEEEpeerreviewmaketitle

	\section{Introduction}\label{S1}
    \IEEEPARstart{T}{he} next-generation wireless communication systems have an increasing demand for network capacity \cite{9174846}.
    The cell-free multi-input multi-output (MIMO) concept constitutes a promising technique \cite{7827017} of enhancing the capacity of wireless communication systems due to its capability of supporting a large number of handsets, by relying on distributed access points (APs) controlled by a central processing unit (CPU) \cite{8845768}.
    The key advantage of the cell-free MIMO architecture is its user-centric (UC) nature, where each user is collaboratively served by all APs without cell boundaries \cite{7827017,8809413,9174860}.
    In fact, typically a number of users form a UC cluster and they are supported by a number of APs.
    When the user density is high, more APs join the cluster with the objective of load balancing, which leads to amorphous clusters of different sizes.
    By cancelling the inter-user interference (IUI) using transmit precoding (TPC), cell-free MIMO significantly improves the network capacity of wireless communications \cite{8113550}.

    Millimeter-wave (mmWave) solutions have been identified as one of the key techniques for future wireless communication systems as a benefit of their abundant bandwidth \cite{8337813,8902153,JTSP_HRobert_OverviewMmwave,Sma}.
    However, the extremely high path-loss of mmWave signals has to be compensated \cite{7109864}.
    Fortunately, thanks to the short wavelength of mmWave carriers, a large number of antennas can be packaged in a compact physical size \cite{8777168,8698333,6894454,5262293}.
    The high spatial gain of a high-dimensional antenna array is able to overcome the severe path-loss \cite{6979962,7010533,6736750}.
    Due to the sparsity of mmWave channels, hybrid TPC has been proposed for achieving high spatial gain despite using a reduced number of radio frequency (RF) chains, which considerably reduces the power consumption of mmWave systems \cite{8284058,7445130,8370683}.
    Therefore, the combination of hybrid TPC and cell-free MIMO has both a high network capacity and a high energy efficiency \cite{2010.09162v1,8676377}.

\begin{table*}[t]
    \centering
  \caption{Boldly Contrasting Our Contributions to the Relevant Literatures}
  \label{tab:Summary}
        \begin{tabular}{| l | c | c | c | c | c | c | c | c |}
        \hline
               & \cite{7827017} & \cite{8845768} & \cite{2010.09162v1} & \cite{8676377} & \cite{7917284} & \cite{2002.03744v2} & \cite{9069486} & \textbf{Proposed}\\
        \hline
        \hline
            Hybrid Precoding        &               &            & \checkmark & \checkmark &            &            &            & \checkmark\\
        \hline
            Uplink                  & \checkmark    & \checkmark & \checkmark & \checkmark &            &            &            &  \\
        \hline
            Downlink                & \checkmark    &            &            & \checkmark & \checkmark & \checkmark & \checkmark &  \checkmark\\
        \hline
            Multiple antennas at each user &        &            & \checkmark &            &            & \checkmark &            & \checkmark\\
        \hline
            Multiple data streams for each user &   &            & \checkmark &            &            &            &            & \checkmark\\
        \hline
            Sum rate                &               & \checkmark & \checkmark & \checkmark &            & \checkmark & \checkmark & \checkmark\\
        \hline
        \end{tabular}
\end{table*}
    \subsection{Literature review}\label{S1.1}
    Some TPC techniques designed for cell-free MIMO networks have been reported in \cite{7827017,8845768,7917284,2002.03744v2,9069486,9110914,8943119}.
    Ngo \textit{et al.} \cite{7827017} propose to use a maximum ratio transmission (MRT)-based TPC at the APs in the downlink (DL) and use a matched filter at the users on the uplink (UL) for maximizing the system throughputs.
    Bj{\"o}rnson and Sanguinetti \cite{8845768} propose to maximize the sum rate of the uplink in cell-free MIMO by using the classical weighted minimum mean square error (WMMSE) based TPC.
    Nayebi \textit{et al.} \cite{7917284} propose to employ a zero-forcing (ZF) TPC on the DL of cell-free MIMO.
    Zhang and Dai \cite{2002.03744v2} propose to maximize the weighted sum rate on the DL of cell-free MIMO by exploiting the recently proposed fractional programming (FP) technique, where only one data stream is considered for each user.
    Furthermore, some prior treatises consider user scheduling in cell-free MIMO.
    Interdonato \textit{et al.} \cite{9069486} propose a pair of distributed TPC schemes in cell-free MIMO, namely the local partial ZF TPC and the local protective partial ZF TPC, which improve the spectral efficiency with the aid of user grouping.
    Nguyen \textit{et al.} \cite{9110914} propose to jointly optimize both the AP-user association and the AP selection for striking a trade-off between a network's spectral efficiency and energy efficiency for cell-free MIMO.
    Wang \textit{et al.} \cite{8943119} consider both a ZF and a regularized ZF (RZF) TPC and advocate a novel genetic algorithm based user scheduling strategy (GAS) for alleviating the uplink-to-downlink interference in cell-free MIMO.
    Moreover, since numerous APs are deployed in cell-free networks, the energy-efficient hybrid precoding structures have substantial benefits in cell-free MIMO \cite{2010.09162v1,8676377}, especially when sparse mmWave channels are considered.

    Hybrid TPCs have been widely studied both in single-cell and multi-cell networks.
    El-Ayach \textit{et al}. \cite{el2013spatially} propose to design a hybrid TPC by using the popular orthogonal matching pursuit (OMP)-based method.
    Alkhateeb and Heath \cite{Alkhateeb2016Frequency} conceive an efficient hybrid analog/digital codebook and propose a codebook-based method for hybrid TPC designs in wideband mmWave systems.
    Sohrabi and Yu \cite{7227015,7389996} conceive an iterative TPC design method for finite resolution of phase shifters, which achieves near-optimal spectral efficiency both in single-user and multi-user MIMO systems.
    Furthermore, some low-complexity analog TPCs have also been proposed.
    Yu \textit{et al}. \cite{8310586} design a switch-based network for dynamically changing the connection between the phase shifters, RF chains and antennas.
    Gao \textit{et al}. \cite{7445130} propose a pioneering sub-array-connected phase-shifter architecture and propose a successive interference cancelation (SIC)-based hybrid TPC method attaining a near-optimal spectral efficiency performance at a low complexity.
    Chen \textit{et al}. \cite{8754694} conceive a randomized two-timescale hybrid TPC scheme for the DL of multicell massive MIMO systems.
    Passive TPC structures have also been proposed by researchers.
    Han \textit{et al.} \cite{8327819} employ the so-called Butler matrices for enhancing discrete fourier transform (DFT)-based systems.
    They also propose a corresponding two-step hybrid TPC method for achieving near-optimal performance at a significantly reduced computational complexity.
    Applying hybrid TPC structures in cell-free MIMO is a promising but challenging field.
The UC property of cell-free MIMO, the power constraints imposed on each AP and the constant-modulus constraints of analog TPC design must be jointly considered. This motivates us to fill in this knowledge-gap.

    There are very few contributions investigating hybrid TPC techniques in cell-free MIMO.
    Femenias \textit{et al}. \cite{8678745} propose to maximize the achievable max-min per-user rate in cell-free MIMO relying on hybrid TPC structures.
    Nguyen \textit{et al}. \cite{2010.09162v1} propose both centralized and semi-centralized hybrid TPC algorithms for maximizing the achievable data rate in the UL of cell-free MIMO.
    However, the derivation of hybrid TPC matrices in the UL of cell-free MIMO cannot be directly applied in the DL.
    Alonzo \textit{et al}. \cite{8676377} propose to decompose the fully digital TPCs into hybrid ones by using a subspace decomposition algorithm for the DL of cell-free MIMO.
    Nevertheless, this subspace decomposition inevitably causes some sum rate loss.
    Therefore, as seen from Table \ref{tab:Summary}, the demand for a general framework of DL hybrid TPC with near-optimal sum rate performance in cell-free MIMO attracts our attention.

    \subsection{Contribution}\label{S1.2}
    Against this background, the main contributions of this paper are summarized as follows:
    \begin{enumerate}
    \item
    We propose to maximize the weighted sum rate (WSR) of the mmWave cell-free MIMO DL relying on hybrid TPC structures, where multiple APs, multiple users and multiple data streams per user are considered.
    We stipulate a pair of constraints for our optimization problem, namely the transmit power constraint of each AP and the constant-modulus constraint imposed on each element of the analog TPC matrices.
    Both the objective function (OF) and the constant-modulus constraint are non-convex, which make the problem formulated difficult to solve.
    To deal with this issue, we first reformulate the OF into an equivalent form by exploiting the WMMSE criterion based upon the UC property of cell-free MIMO.
    Furthermore, we introduce auxiliary TPC matrices and propose a low-complexity block coordinate descent (BCD) algorithm for iteratively designing near-optimal hybrid TPC matrices.
    \item
    In each iteration of the proposed BCD algorithm, we exploit the classic Lagrangian multiplier method combined with the penalty dual decomposition (PDD) method for designing the hybrid analog/digital TPC matrices of the cell-free MIMO DL.
    Specifically, we first derive the optimal closed-form Lagrangian multiplier-based solutions for finding the auxiliary TPC matrices through sophisticated matrix manipulations and transformations.
    Subsequently, we optimize the penalty items by adopting the classic least square (LS) method, which minimizes the Euclidian distance between the auxiliary TPC matrices and the hybrid analog/digital TPC matrices.
    \item
    We further extend our proposed BCD algorithm to the fully digital cell-free MIMO DL.
    In contrast to the hybrid TPC scenario, there is no penalty item in the fully digital TPC design, which leads to the difficulties of low-rank matrix inversion for obtaining the optimal Lagrangian multipliers.
    Therefore, we circumvent this problem by deriving the exact ranks of the key intermediate matrices, which are used for calculating Lagrangian multipliers.
    Our simulation results show that the proposed algorithm can achieve better WSR performance than its conventional counterparts.
    \end{enumerate}

    The remainder of this paper is organized as follows.
    In Section \ref{S2}, our system model, channel model and problem formulation are introduced.
    In Section \ref{S3}, our BCD algorithm proposed for mmWave cell-free MIMO hybrid TPC design is presented.
    In Section \ref{S4}, we extend our proposed BCD algorithm to the mmWave cell-free MIMO DL relying on a fully digital TPC structure.
    In Section \ref{S5}, our numerical results are provided.
    Finally, our conclusions are drawn in Section \ref{S6}.

    \emph{Notation}:
	Lower-case and upper-case boldface letters denote vectors and matrices, respectively;
	$(\cdot)^{\rm{T}}$, $(\cdot)^{\text{H}}$, $(\cdot)^{-1}$ and $(\cdot)^{\dagger}$ denote the transpose, conjugate transpose, inverse and pseudo-inverse of a matrix, respectively;
    $\otimes$ denotes the  Kronecker product;
    $\mathrm{Tr}(\cdot)$ presents the trace function;
    $\| \cdot \|_{F}$ denotes the Frobenius norm of a matrix;
    $|a|$ is the absolute value of a scalar;
    $|\mathbf{A}|$ is the determinant of a matrix;
    $\mathbf{A}_{\left[i,:\right]}$ and $\mathbf{A}_{\left[:,j\right]}$ represent the $i$-th row and $j$-th column of the matrix $\mathbf{A}$, respectively;
    $\mathbf{A}_{\left[a:b,:\right]}$ and $\mathbf{A}_{\left[:,c:d\right]}$ represent the sub-matrices of the matrix $\mathbf{A}$ containing the $a$-th to the $b$-th rows and the $c$-th to the $d$-th columns, respectively;
	Finally, $\mathbf{I}_P$ denotes the identity matrix of size $P\times P$.
	\section{System Model and Channel Model of Cell-Free MIMO Relying on Hybrid Precoding Structures}\label{S2}
    In this section, we introduce the narrow-band mmWave cell-free MIMO DL model, where hybrid TPCs are used at the APs.
	\subsection{System Model of Cell-Free MIMO Relying on Hybrid Precoding Structures}\label{S2.1}
 	\begin{figure}[t]
		\center{\includegraphics[width=0.5\textwidth]{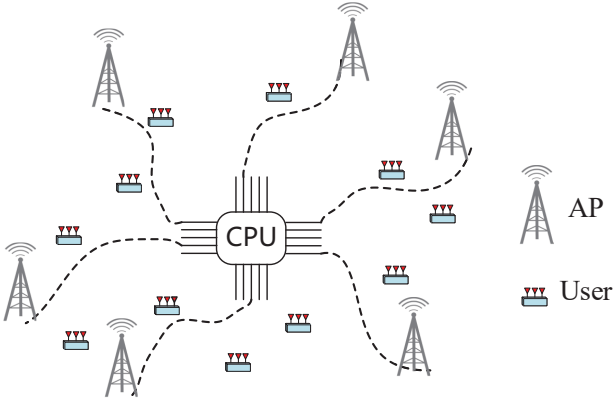}}
		\caption{Illustration of the cell-free MIMO concept.}
		\label{cell_free}
	\end{figure}
 	\begin{figure}[t]
		\center{\includegraphics[width=0.45\textwidth]{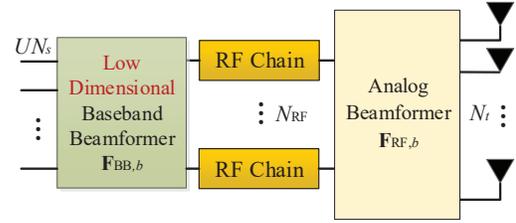}}
		\caption{Illustration of the hybrid TPC structure at each AP.}
		\label{HB_structure}
	\end{figure}
    As shown in Fig. \ref{cell_free}, there are $B$ multiple-antenna APs relying on hybrid TPCs for supporting $U$ multiple-antenna users in the cell-free MIMO network.
    Each AP is equipped with $N_t$ transmit antennas (TAs) and $N_{\rm RF}$ RF chains, $N_{\rm RF} \leq N_t $, as shown in Fig. \ref{HB_structure}.
    At each user, fully digital structures are adopted, where $N_r$ receive antennas (RAs) are employed for receiving $N_s$ data streams, $N_s \leq N_r$.
    We assume that $UN_s \leq BN_{\rm RF}$ in our cell-free MIMO.
    For simplicity, we define $\mathcal{N}_t = \left\{1,2,\cdots, N_t\right\}$, $\mathcal{N}_r = \left\{1,2,\cdots, N_r\right\}$, $\mathcal{N}_{\rm RF} = \left\{1,2,\cdots, N_{\rm RF}\right\}$, $\mathcal{N}_s = \left\{1,2,\cdots, N_s\right\}$, $\mathcal{U} = \left\{1,2,\cdots, U\right\}$, $\mathcal{B} = \left\{1,2,\cdots, B\right\}$.

    We assume that all APs serve all users. Thus, all APs synchronously transmit data symbol $\left\{\mathbf{s}_u\right\}_{u=1}^U$, where $\mathbf{s}_u \in \mathbb{C}^{N_s \times 1}$ is the data symbols intended for the $u$-th user.
    The transmitted data symbols are assumed to have normalized power, i.e., $\mathbb{E}\left\{ \mathbf{s}_u \mathbf{s}_u^{\rm H} \right\} = \mathbf{I}_{N_s} $, $\forall u \in \mathcal{U}$.
    The data symbols $\mathbf{s}_u$ transmitted from the $b$-th AP are first digitally precoded by $\mathbf{F}_{{\rm BB},b,u} \in \mathbb{C}^{N_{\rm RF} \times N_s}$.
    Then the analog TPC $\mathbf{F}_{{\rm RF},b} \in \mathbb{C}^{N_t \times N_{\rm RF}}$ relying on phase shifters is applied.
    Therefore, the signals $\mathbf{x}_b \in \mathbb{C}^{N_t \times 1}$ after hybrid precoding at the $b$-th AP are expressed by
    \begin{align}\label{ts_b}
    \mathbf{x}_{b} = \mathbf{F}_{{\rm RF},b}\sum_{u=1}^U \mathbf{F}_{{\rm BB},b,u}\mathbf{s}_{u}.
	\end{align}

    We define the mmWave channel between the $b$-th AP and the $u$-th user as $\mathbf{H}_{b,u} \in \mathbb{C}^{N_{r} \times N_{t}}$.
    It is assumed that the CSI is perfectly known at all APs \cite{9154244,8741198,9110869}, noting that both the accurate channel estimation and the robust TPC design relying on partial CSI constitute rather different problems in cell-free MIMO systems \cite{alexandropoulos2017robustness,9366805,9136592}. Given the strict length-limitation, these have to be set aside for future research.
    Then the signals $\mathbf{y}_u \in \mathbb{C}^{N_r \times 1}$ received at the $u$-th user are expressed by
    \begin{align}\label{y_uk2}
    \mathbf{y}_{u} = \sum_{b=1}^B\mathbf{H}_{b,u}\mathbf{x}_{b} & = \underbrace{\sum_{b=1}^B\mathbf{H}_{b,u}\mathbf{F}_{{\rm RF},b}\mathbf{F}_{{\rm BB},b,u}\mathbf{s}_{u}}_{\text{desired signal of the $u$-th user}}\nonumber \\
    & + \underbrace{\sum_{b=1}^B\sum_{j=1,j\neq u}^U\mathbf{H}_{b,u}\mathbf{F}_{{\rm RF},b}\mathbf{F}_{{\rm BB},b,j}\mathbf{s}_{j}}_{\text{IUI}} + \mathbf{n}_{u},
	\end{align}
    where the first item represents the desired signal of the $u$-th user, the second item stands for the IUI, and $\mathbf{n}_{u}\sim \mathcal{CN}\left(0, \sigma^2\mathbf{I}_{N_r} \right) \in \mathbb{C}^{N_r\times 1}$ is the additive white Gaussian noise (AWGN).
    We further reformulate (\ref{y_uk2}) as
    \begin{align}\label{y_uk3}
    \mathbf{y}_{u} = \mathbf{H}_{u}\mathbf{F}_{{\rm HB},u}\mathbf{s}_{u} + \sum_{j=1,j\neq u}^U\mathbf{H}_{u}\mathbf{F}_{{\rm HB},j}\mathbf{s}_{j} + \mathbf{n}_{u},
	\end{align}
    where $\mathbf{H}_{u} = \left[\mathbf{H}_{1,u},\mathbf{H}_{2,u}, \cdots, \mathbf{H}_{B,u}\right] \in \mathbb{C}^{N_{r} \times BN_t} $, $\mathbf{F}_{{\rm HB},u} = \left[\mathbf{F}_{{\rm HB},1,u}^{\rm H}, \mathbf{F}_{{\rm HB},2,u}^{\rm H}, \cdots, \mathbf{F}_{{\rm HB},B,u}^{\rm H} \right]^{\rm H}\in \mathbb{C}^{BN_t \times N_s}$, and $\mathbf{F}_{{\rm HB},b,u} = \mathbf{F}_{{\rm RF},b}\mathbf{F}_{{\rm BB},b,u}\in \mathbb{C}^{N_t \times N_s}$.
    \subsection{Narrow-band MmWave Channel Model}\label{S2.2}
In this paper, we adopt the popular physical multi-path mmWave channel model, where uniform planar arrays (UPAs) of antennas are employed both at APs and users \cite{el2013spatially,7389996,8331836}.
Therefore, we can express the channel $\mathbf{H}_{b,u}$ between the $b$-th AP and the $u$-th user by
    \begin{align}\label{H}
    \mathbf{H}_{b,u} = \sqrt{\frac{N_t N_r}{L}} \sum_{\ell = 1}^{L}\alpha_{\ell,b,u}\mathbf{a}_{r}\left( \phi^{\mathrm{r}}_{\ell,b,u}, \theta^{\mathrm{r}}_{\ell,b,u} \right)\mathbf{a}_{t}^{\text{H}}\left( \phi^{\mathrm{t}}_{\ell,b,u}, \theta^{\mathrm{t}}_{\ell,b,u} \right),
	\end{align}
where $L$ denotes the number of paths, $\alpha_{\ell,b,u}\sim \mathcal{CN}\left(0, 1\right)$ is the complex gain of the $\ell$-th path between the $b$-th AP and the $u$-th user.
The vectors $\mathbf{a}_{t}\left( \phi^{\mathrm{t}}_{\ell,b,u}, \theta^{\mathrm{t}}_{\ell,b,u} \right) \in \mathbb{C}^{N_{t} \times 1 }$ and $\mathbf{a}_{r}\left( \phi^{\mathrm{r}}_{\ell,b,u}, \theta^{\mathrm{r}}_{\ell,b,u} \right)\in \mathbb{C}^{N_{r} \times 1 }$ represent the antenna array responses at the $b$-th AP and the $u$-th user, which can be expressed as
    \begin{align}\label{a_BS}
    &\mathbf{a}_{t}\left( \phi^{\mathrm{t}}_{\ell,b,u}, \theta^{\mathrm{t}}_{\ell,b,u} \right) \nonumber \\
    & = \frac{1}{\sqrt{N_{t}}}  \left[ 1,\cdots, e^{j\frac{2\pi}{\lambda}d\left( m \mathrm{sin} \phi^{\mathrm{t}}_{\ell,b,u} \mathrm{sin} \theta^{\mathrm{t}}_{\ell,b,u} + n \mathrm{cos}\theta^{\mathrm{t}}_{\ell,b,u} \right) } , \right.\nonumber\\
    & \cdots, \left.e^{j\frac{2\pi}{\lambda}d\left( \left( W-1 \right) \mathrm{sin} \phi^{\mathrm{t}}_{\ell,b,u} \mathrm{sin} \theta^{\mathrm{t}}_{\ell,b,u} + \left( H-1 \right) \mathrm{cos}\theta^{\mathrm{t}}_{\ell,b,u} \right) } \right]^{\text{T}},
	\end{align}
where $\phi^{\mathrm{t}}_{\ell,b,u}$ and $\theta^{\mathrm{t}}_{\ell,b,u}$ denote the azimuth and elevation angle-of-departure (AoD) from the $b$-th AP to the $u$-th user, $\lambda$ represents the wavelength, and $d$ is the antenna spacing.
Furthermore, $m \in \{0,1,\cdots,W-1 \}$ and $n \in \{0,1,\cdots,H-1 \}$ with $W$ and $H$ denoting the number of antennas in horizontal and vertical directions.
Similarly, $\mathbf{a}_{r}\left( \phi^{\mathrm{r}}_{\ell,b,u}, \theta^{\mathrm{r}}_{\ell,b,u} \right)$ can be expressed in the same form as (\ref{a_BS}) by replacing $\phi^{\mathrm{t}}_{\ell,b,u}$ and $\theta^{\mathrm{t}}_{\ell,b,u}$ with the azimuth and elevation angle-of-arrival (AoA) $\phi^{\mathrm{r}}_{\ell,b,u}$ and $\theta^{\mathrm{r}}_{\ell,b,u}$.
    \subsection{Problem Formulation}\label{S2.3}
    Again, we aim for maximizing the WSR of the mmWave cell-free MIMO DL by designing the hybrid TPC matrices of all $B$ APs.
    We first present the achievable WSR expression of
    \begin{align}\label{SumRate}
    R = \sum_{u=1}^{U} \omega_{u} \log \left| \mathbf{I}_{N_r} + \mathbf{J}^{-1}_{u}\mathbf{H}_{u}\mathbf{F}_{{\rm HB},u}\mathbf{F}_{{\rm HB},u}^{\mathrm{H}}\mathbf{H}_{u}^{\mathrm{H}} \right|,
	\end{align}
    where $\mathbf{J}_{u} = \sum_{j=1,j\neq u}^U\mathbf{H}_{u}\mathbf{F}_{{\rm HB},j}\mathbf{F}_{{\rm HB},j}^{\rm H}\mathbf{H}_{u}^{\rm H} + \sigma^2\mathbf{I}_{N_r}$ denotes the interference-plus-noise item, and $\omega_u$ is the weighting coefficient of the $u$-th user.
    The WSR maximization problem of the mmWave cell-free MIMO DL relying on hybrid TPC is formulated as
    \begin{subequations}\label{ObjEE_v1}
    \begin{align}
    \max_{ \mathcal{F}_{\rm RF}, \mathcal{F}_{\rm BB} } \quad &  R \\
    s.t. \quad \ &  \sum_{u=1}^U \mathrm{Tr}\left( \mathbf{F}_{{\rm RF},b}\mathbf{F}_{{\rm BB},b,u}\mathbf{F}_{{\rm BB},b,u}^{\rm H}\mathbf{F}_{{\rm RF},b}^{\rm H} \right) \leq P_{b,\max}, \nonumber \\
    & \quad\quad\quad\quad\quad\quad\quad\quad \quad\quad\quad\quad\quad\quad \forall b \in \mathcal{B},  \\
    &   \left|  \mathbf{F}_{{\rm RF},b}\left( i,j \right) \right| = 1, \forall i \in \mathcal{N}_t, \forall j \in \mathcal{N}_{\rm RF}, \forall b \in \mathcal{B},
    \end{align}
    \end{subequations}
    where $ \mathcal{F}_{\rm RF} = \left\{ \mathbf{F}_{{\rm RF},b} | \forall b \in \mathcal{B}\right\}$, $ \mathcal{F}_{\rm BB} = \left\{ \mathbf{F}_{{\rm BB},b,u} | \forall b \in \mathcal{B},\forall u \in \mathcal{U}\right\}$.
    The constraint of (\ref{ObjEE_v1}b) ensures that the transmit power at the $b$-th AP is lower than the maximum power $P_{b,\max}$.
    The constraint of  (\ref{ObjEE_v1}c) is due to the constant-modulus constraint of phase shifters used in the analog TPCs.
    We observe that both of the OF (\ref{ObjEE_v1}a) and the constraint (\ref{ObjEE_v1}c) are non-convex.
    Hence it is challenging to solve the problem (\ref{ObjEE_v1}) directly.
    \section{Proposed Hybrid Precoding Algorithm for MmWave cell-free MIMO}\label{S3}
    In this section, we propose a low complexity method for solving the problem (\ref{ObjEE_v1}).
    Firstly, we transform the problem into an equivalent form by exploiting the equivalence between the sum rate and the WMMSE \cite{7890495,5756489,alexandropoulos2013reconfigurable,7509379}.
    Then we propose a BCD algorithm for solving the problem in an alternating optimization manner.
    The analog and digital TPCs are alternately optimized by applying the Lagrangian multiplier method combined with the PDD method, which is inspired by \cite{9120361,9119203,9110865}.
    \subsection{Problem Transformation}\label{S3.1}
    On basis of the UC nature of the cell-free MIMO DL, the problem (\ref{ObjEE_v1}) can be equivalently transformed into a WMMSE problem by introducing a set of linear combiners $\mathcal{G} = \left\{ \mathbf{G}_u  \in \mathbb{C}^{N_r\times N_s} | \forall u \in \mathcal{U} \right\}$ and a set of arbitrary positive definite weighting matrices $\mathcal{W} = \left\{ \mathbf{W}_u  \in \right.$ $ \left.\mathbb{C}^{N_s\times N_s} | \forall u \in \mathcal{U} \right\}$.
    Specifically, the signals $\hat{\mathbf{s}}_{u}  \in \mathbb{C}^{N_s\times 1}$ estimated by the $u$-th user acquired with the aid of $\mathbf{G}_u$ are expressed as
    \begin{align}\label{s_esti}
    \hat{\mathbf{s}}_{u} =\mathbf{G}_{u}^{\text{H}}\mathbf{y}_{u}.
	\end{align}
    Then, we are ready to derive the MSE covariance matrix for the $u$-th user as
    \begin{align}\label{MSE}
    \mathbf{E}_{u} & = \mathbb{E}\left[ \left( \hat{\mathbf{s}}_{u} - \mathbf{s}_{u} \right)\left( \hat{\mathbf{s}}_{u} - \mathbf{s}_{u} \right)^{\text{H}} \right]  \nonumber \\
    & = \left( \mathbf{G}_{u}^{\text{H}}\mathbf{H}_{u}\mathbf{F}_{{\rm HB},u} - \mathbf{I}_{N_s} \right)\left( \mathbf{G}_{u}^{\text{H}}\mathbf{H}_{u}\mathbf{F}_{{\rm HB},u} - \mathbf{I}_{N_s} \right)^{\text{H}} \nonumber \\
    & +  \mathbf{G}_{u}^{\text{H}} \mathbf{J}_{u} \mathbf{G}_{u}.
	\end{align}
    The problem (\ref{ObjEE_v1}) can be equivalently transformed into the WMMSE problem, as detailed in \cite{7890495,5756489,alexandropoulos2013reconfigurable,7509379}
    \begin{subequations}\label{WMMSE}
    \begin{align}
    & \max_{ \mathcal{W},\mathcal{G},\mathcal{F}_{\rm RF}, \mathcal{F}_{\rm BB} } \  \sum_{u=1}^U \omega_u \left[ \log\left| \mathbf{W}_{u} \right| - \mathrm{Tr}\left( \mathbf{W}_{u}\mathbf{E}_{u} \right) + \mathrm{Tr}\left( \mathbf{I}_{N_s} \right) \right] \\
    & \quad \quad s.t.  \quad \quad \quad \text{(\ref{ObjEE_v1}b)} , \quad \text{(\ref{ObjEE_v1}c)}.
    \end{align}
    \end{subequations}
    However, the problem (\ref{WMMSE}) is still challenging to solve due to the non-convex constraint of (\ref{ObjEE_v1}c).
    To deal with this problem, we introduce a set of auxiliary TPC matrices $\mathcal{F} = \left\{ \mathbf{F}_u \in \mathbb{C}^{BN_t \times N_s}| \forall u \in \mathcal{U}\right\}$ as the approximate solution of $\left\{ \mathbf{F}_{{\rm HB},u} | \forall u \in \mathcal{U}\right\}$.
    Then, relying on the PDD method \cite{9120361,9119203}, we introduce a set of penalty parameters $\left\{\rho_b | \forall b \in \mathcal{B}\right\}$ and the corresponding penalty terms $\frac{1}{2\rho_b}\left\| \mathbf{F}_{b,u} - \mathbf{F}_{{\rm HB},b,u} \right\|_F^2$, where $\mathbf{F}_{b,u} = \mathbf{F}_{u\left[ \left( b-1 \right)N_t+1:bN_t ,: \right]}$ denotes the submatrix of $\mathbf{F}_{u}$.
    By neglecting the constant term $\mathrm{Tr}\left( \mathbf{I}_{N_s} \right)$, the problem (\ref{WMMSE}) is transformed into
    \begin{subequations}\label{PDD}
    \begin{align}
    & \max_{ \mathcal{W},\mathcal{G},\mathcal{F},\mathcal{F}_{\rm RF}, \mathcal{F}_{\rm BB} } \  \sum_{u=1}^U \omega_u \left[ \log\left| \mathbf{W}_{u} \right| - \mathrm{Tr}\left( \mathbf{W}_{u}\widetilde{\mathbf{E}}_{u} \right) \right] \nonumber \\
    & \quad \quad \quad \quad \quad \quad \quad  - \sum_{b=1}^B\sum_{u=1}^U\frac{1}{2\rho_b}\left\| \mathbf{F}_{b,u} - \mathbf{F}_{{\rm HB},b,u} \right\|_F^2 \\
    & \quad \quad s.t. \quad \sum_{u=1}^U \mathrm{Tr}\left( \mathbf{F}_{b,u}\mathbf{F}_{,b,u}^{\rm H}\right) \leq P_{b,\max}, \forall b \in \mathcal{B}, \quad \text{(\ref{ObjEE_v1}c)}
    \end{align}
    \end{subequations}
    where we have
    \begin{align}\label{tilde_MSE}
    \widetilde{\mathbf{E}}_{u} & =  \left( \mathbf{G}_{u}^{\text{H}}\mathbf{H}_{u}\mathbf{F}_{u} - \mathbf{I}_{N_s} \right)\left( \mathbf{G}_{u}^{\text{H}}\mathbf{H}_{u}\mathbf{F}_{u} - \mathbf{I}_{N_s} \right)^{\text{H}} \nonumber \\
    & + \sum_{j=1,j\neq u}^U\mathbf{G}_{u}^{\text{H}} \mathbf{H}_{u}\mathbf{F}_{j}\mathbf{F}_{j}^{\rm H}\mathbf{H}_{u}^{\rm H}\mathbf{G}_{u} + \sigma^2 \mathbf{G}_{u}^{\text{H}} \mathbf{G}_{u}.
	\end{align}
    We observe that the problem (\ref{PDD}) is more tractable because the OF (\ref{PDD}a) is a concave function with respect to $\mathcal{W}$,  $\mathcal{G}$, and $\mathcal{F}$, provided that the other matrices are fixed.
    Hence the problem (\ref{PDD}) is a convex optimization problem.
    Moreover, the hybrid analog/digital TPC matrices $\mathcal{F}_{\rm RF}$ and $\mathcal{F}_{\rm BB}$ are only related to the penalty terms in (\ref{PDD}a).
    Therefore, we can derive the closed-form solution of $\mathcal{W}$,  $\mathcal{G}$, $\mathcal{F}$, $\mathcal{F}_{\rm RF}$ and $\mathcal{F}_{\rm BB}$ by exploiting the BCD algorithm.
    Briefly, the key idea of the BCD algorithm is to iteratively obtain one set of variables while keeping the others fixed.
    \subsection{Optimization of $\mathcal{G}$ and $\mathcal{W}$}\label{S3.2}
    The closed-form solutions of $\mathcal{G}$ and $\mathcal{W}$ can be acquired by using the partial derivative method.
    Specifically, after substituting (\ref{tilde_MSE}) into (\ref{WMMSE}a), we derive the partial derivative of (\ref{WMMSE}a) with respect to $\mathbf{G}_u$ and set it to zero.
    Then the optimal solution of $\mathbf{G}_u$ for the $u$-th user is given by
    \begin{align}\label{G_uopt}
    \mathbf{G}_{u}^{\star} =  \left( \sum_{j=1}^U \mathbf{H}_{u}\mathbf{F}_{j}\mathbf{F}_{j}^{\rm H}\mathbf{H}_{u}^{\rm H} + \sigma^2\mathbf{I}_{N_r} \right)^{-1} \mathbf{H}_u \mathbf{F}_u.
	\end{align}
    Similarly, we can derive the optimal solution of $\mathbf{W}_u$ for the $u$-th user as
    \begin{align}\label{W_uopt}
    \mathbf{W}_{u}^{\star} = \widetilde{\mathbf{E}}_{u}^{-1}.
	\end{align}
    \subsection{Optimization of Approximate Hybrid Precoding Matrices $\mathcal{F}$}\label{S3.3}
    Given both the linear combining matrices $\mathcal{G}$, as well as the weighting matrices $\mathcal{W}$, and the analog TPC matrices $\mathcal{F}_{\rm RF}$ and the digital TPC matrices $\mathcal{F}_{\rm BB}$, the subproblem with respect to $\mathcal{F}$ is expressed as
    \begin{subequations}\label{sub_probl_F}
    \begin{align}
    & \min_{\mathcal{F}} \  \sum_{u=1}^U \omega_u \left[\mathrm{Tr}\left( \mathbf{W}_{u}\widetilde{\mathbf{E}}_{u} \right) \right] \nonumber \\
    & \quad \quad\quad + \sum_{b=1}^B\sum_{u=1}^U\frac{1}{2\rho_b}\left\| \mathbf{F}_{b,u} - \mathbf{F}_{{\rm HB},b,u} \right\|_F^2 \\
    & \ s.t. \quad \sum_{u=1}^U \mathrm{Tr}\left( \mathbf{F}_{b,u}\mathbf{F}_{,b,u}^{\rm H}\right) \leq P_{b,\max}, \forall b \in \mathcal{B},
    \end{align}
    \end{subequations}
    which is transformed from the problem (\ref{PDD}) by neglecting the variables irrelevant to $\mathcal{F}$.
    We can derive the closed-form solutions of the problem (\ref{sub_probl_F}) by applying the Lagrangian multiplier method.
    Firstly, we transform the problem (\ref{sub_probl_F}) into an unconstrained form by introducing a set of Lagrangian multipliers $\left\{ \lambda_b \geq 0 | \forall b \in \mathcal{B} \right\}$ as
    \begin{align}\label{sub_probl_F_Lagrangian}
    \min_{\left\{ \lambda_b | \forall b \in \mathcal{B} \right\},\mathcal{F}} \ & \sum_{u=1}^U\omega_u\left[\mathrm{Tr}\left( \mathbf{W}_{u}\widetilde{\mathbf{E}}_{u} \right) \right] \nonumber \\
    & \quad\quad\quad+ \sum_{b=1}^B\sum_{u=1}^U\frac{1}{2\rho_b}\left\| \mathbf{F}_{b,u} - \mathbf{F}_{{\rm HB},b,u} \right\|_F^2 \nonumber \\
    & + \sum_{b=1}^B \lambda_b \left[ \sum_{u=1}^U \mathrm{Tr}\left( \mathbf{F}_{b,u}\mathbf{F}_{,b,u}^{\rm H}\right) - P_{b,\max} \right].
    \end{align}
    By substituting (\ref{tilde_MSE}) into (\ref{sub_probl_F_Lagrangian}), we further transform the problem (\ref{sub_probl_F_Lagrangian}) as follows
    \begin{align}\label{sub_probl_F_Lagrangian_v2}
    \min_{\left\{ \lambda_b | \forall b \in \mathcal{B} \right\},\mathcal{F}} \ & \sum_{u=1}^U\omega_u\left[\mathrm{Tr}\left( \mathbf{F}_{u}^{\rm H} \mathbf{A}_u \mathbf{F}_{u} \right)\right] \nonumber \\
    & - \sum_{u=1}^U\omega_u\left[\mathrm{Tr}\left( \mathbf{F}_{u}^{\rm H}\mathbf{H}_{u}^{\rm H}\mathbf{G}_{u}\mathbf{W}_{u} \right) \right] \nonumber \\
    & - \sum_{u=1}^U\omega_u\left[\mathrm{Tr}\left( \mathbf{W}_{u}\mathbf{G}_{u}^{\rm H}\mathbf{H}_{u}\mathbf{F}_{u} \right)\right] \nonumber\\
    & + \sum_{b=1}^B\sum_{u=1}^U\frac{1}{2\rho_b}\left\| \mathbf{F}_{b,u} - \mathbf{F}_{{\rm HB},b,u} \right\|_F^2 \nonumber \\
    & + \sum_{b=1}^B \lambda_b \left[ \sum_{u=1}^U \mathrm{Tr}\left( \mathbf{F}_{b,u}\mathbf{F}_{,b,u}^{\rm H}\right) - P_{b,\max} \right],
    \end{align}
    where $\mathbf{A}_u = \mathbf{H}_u^{\rm H}\mathbf{G}_u\mathbf{W}_u\mathbf{G}_u^{\rm H}\mathbf{H}_u \in \mathbb{C}^{BN_t \times BN_t}$.
    The derivation of the closed-form $\mathbf{F}_{u}$ and $\left\{ \lambda_b \right\}_{b=1}^B$ is more elaborate in our mmWave cell-free MIMO scenario, since the last two items of (\ref{sub_probl_F_Lagrangian_v2}) are functions of $\mathbf{F}_{b,u}$, which are the submatrices of $\mathbf{F}_{u}$.
    These two items are introduced for satisfying the transmit power constraints of each AP and the constant-modulus constrains of phase shifters.
    These two constrains of our mmWave cell-free MIMO DL relying on hybrid TPC structures differentiate our problem formulation from that of the existing literatures \cite{9110849,9090356}.
    For solving this challenging problem, we have the following proposition.

    \textbf{Proposition 1}:
    We propose to iteratively update $\left\{ \mathbf{F}_{b,u} | \forall b \in \mathcal{B}, \forall u \in \mathcal{U} \right\}$ in each iteration of the BCD algorithm.
    Hence we recast problem (\ref{sub_probl_F_Lagrangian_v2}) first.
    Specifically, we define $\mathbf{A}_u = \mathbf{A}_u^{\frac{1}{2}} \left( \mathbf{A}_u^{\frac{1}{2}} \right)^{\rm H}$ and $\widetilde{\mathbf{A}}_{b,u} = \widehat{\mathbf{A}}_{b,u}\sum_{i=1,i\neq b}^{B}\widehat{\mathbf{A}}_{b,i}^{\rm H}\mathbf{F}_{b,i} \in \mathbb{C}^{N_t \times N_s}$, where $\left\{\widehat{\mathbf{A}}_{b,u} \in \mathbb{C}^{N_t \times BN_t} | \forall b \in \mathcal{B} \right\}$ are a set of variables introduced for constructing $\mathbf{A}_{u}^{\frac{1}{2}} = \left[ \widehat{\mathbf{A}}_{1,u}^{\rm H}, \right.$ $\left.\widehat{\mathbf{A}}_{2,u}^{\rm H}, \cdots, \widehat{\mathbf{A}}_{B,u}^{\rm H} \right]^{\rm H}$.
    We also introduce $\left\{\mathbf{C}_{b,u} \in \mathbb{C}^{N_t \times N_s} | \forall b \in \mathcal{B} \right\}$ for constructing $\mathbf{C}_{u} = \omega_u\mathbf{H}_{u}^{\rm H}\mathbf{G}_{u}\mathbf{W}_{u} = \left[ \mathbf{C}_{1,u}^{\rm H}, \mathbf{C}_{2,u}^{\rm H}, \cdots, \mathbf{C}_{B,u}^{\rm H} \right]^{\rm H}$.
    Then the problem (\ref{sub_probl_F_Lagrangian_v2}) can be characterized as
    \begin{align}\label{sub_probl_F_Lagrangian_v3}
    & \min_{\left\{ \mathbf{F}_{b,u} | \forall b \in \mathcal{B}, \forall u \in \mathcal{U} \right\}} \nonumber\\
    &\sum_{u=1}^U\left[\mathrm{Tr}\left( \mathbf{F}_{b,u}^{\rm H} \left(\mathbf{Q}_{b,u} + \frac{1}{2\rho_b}\mathbf{I}_{N_t} + \lambda_b \mathbf{I}_{N_t} \right) \mathbf{F}_{b,u} \right)\right] \nonumber \\
    & + \sum_{i=1,i\neq b}^B\sum_{u=1}^U\left[\mathrm{Tr}\left( \mathbf{F}_{i,u}^{\rm H} \left(\frac{1}{2\rho_b}\mathbf{I}_{N_t} + \lambda_b \mathbf{I}_{N_t} \right) \mathbf{F}_{i,u} \right)\right] \nonumber \\
    & - \sum_{u=1}^U\left[\mathrm{Tr}\left( \left(\mathbf{M}_{b,u}  + \frac{1}{2\rho_b}\mathbf{F}_{{\rm HB},b,u}\right)^{\rm H}\mathbf{F}_{b,u} \right)\right]\nonumber \\
    & - \sum_{i=1,i\neq b}^B\sum_{u=1}^U\left[\mathrm{Tr}\left( \left(\mathbf{C}_{i,u}  + \frac{1}{2\rho_b}\mathbf{F}_{{\rm HB},i,u}\right)^{\rm H}\mathbf{F}_{i,u} \right)\right]\nonumber \\
    & - \sum_{u=1}^U\left[\mathrm{Tr}\left( \mathbf{F}_{b,u}^{\rm H}\left(\mathbf{M}_{b,u}  + \frac{1}{2\rho_b}\mathbf{F}_{{\rm HB},b,u}\right) \right) \right]\nonumber \\
    & - \sum_{i=1,i\neq b}^B\sum_{u=1}^U\left[\mathrm{Tr}\left( \mathbf{F}_{i,u}^{\rm H}\left(\mathbf{C}_{i,u}  + \frac{1}{2\rho_b}\mathbf{F}_{{\rm HB},i,u}\right) \right) \right].
    \end{align}
    where we have $\mathbf{Q}_{b,u} = \omega_u\widehat{\mathbf{A}}_{b,u} \widehat{\mathbf{A}}_{b,u}^{\rm H}$ and $\mathbf{M}_{b,u} = \mathbf{C}_{b,u} - \omega_u\widetilde{\mathbf{A}}_{b,u}$.
    \begin{IEEEproof}
        Please see Appendix A.
    \end{IEEEproof}
    Then the optimal $\left\{ \mathbf{F}_{b,u} | \forall b \in \mathcal{B}, \forall u \in \mathcal{U} \right\}$ can be determined by deriving the partial derivative of (\ref{sub_probl_F_Lagrangian_v3}) with respect to $\mathbf{F}_{b,u}$ and setting it to zero.
    Since $\mathbf{F}_{b,u}$ is a function of $\lambda_b$, the optimal $\mathbf{F}_{b,u}\left( \lambda_b \right)$ is expressed as
    \begin{align}\label{F_buopt}
    &\mathbf{F}_{b,u}^{\star}\left( \lambda_b \right) \nonumber \\
    & = \left(\mathbf{Q}_{b,u} + \frac{1}{2\rho_b}\mathbf{I}_{N_t} + \lambda_b \mathbf{I}_{N_t} \right)^{-1} \left(\mathbf{M}_{b,u}  + \frac{1}{2\rho_b}\mathbf{F}_{{\rm HB},b,u}\right).
	\end{align}
    Let us now discuss the derivation of the Lagrangian multipliers $\left\{ \lambda_b | \forall b \in \mathcal{B} \right\}$.
    By substituting (\ref{F_buopt}) into the power constraints of (\ref{sub_probl_F}b), we arrive at the following transformation
    \begin{align} \label{a_trans1}
    & \mathrm{Tr}\left( \left( \mathbf{F}_{b,u}^{\star}\left( \lambda_b \right) \right)^{\rm H}  \mathbf{F}_{b,u}^{\star}\left( \lambda_b \right) \right) \nonumber \\
    = & \mathrm{Tr}\left(  \left(\mathbf{M}_{b,u}  + \frac{1}{2\rho_b}\mathbf{F}_{{\rm HB},b,u}\right)^{\rm H} \right. \nonumber \\
    & \left.\left( \left(\mathbf{Q}_{b,u} + \frac{1}{2\rho_b}\mathbf{I}_{N_t} + \lambda_b \mathbf{I}_{N_t}\right)^{\rm H} \right)^{-1} \right. \nonumber \\
    & \left. \cdot \left(\mathbf{Q}_{b,u} + \frac{1}{2\rho_b}\mathbf{I}_{N_t} + \lambda_b \mathbf{I}_{N_t} \right)^{-1} \left(\mathbf{M}_{b,u}  + \frac{1}{2\rho_b}\mathbf{F}_{{\rm HB},b,u}\right) \right) \nonumber \\
    \overset{\left(a\right)}{=} &\mathrm{Tr}\left(  \left(\mathbf{M}_{b,u}  + \frac{1}{2\rho_b}\mathbf{F}_{{\rm HB},b,u}\right)^{\rm H}\right. \nonumber \\
    & \left. \left( \left(\mathbf{U}_{b,u}\mathbf{\Sigma}_{b,u}\mathbf{U}_{b,u}^{\rm H} + \frac{1}{2\rho_b}\mathbf{I}_{N_t} + \lambda_b \mathbf{I}_{N_t}\right)^{\rm H} \right)^{-1} \right. \nonumber \\
    & \left. \cdot \left(\mathbf{U}_{b,u}\mathbf{\Sigma}_{b,u}\mathbf{U}_{b,u}^{\rm H} + \frac{1}{2\rho_b}\mathbf{I}_{N_t} + \lambda_b \mathbf{I}_{N_t} \right)^{-1}\right. \nonumber \\
    & \left. \left(\mathbf{M}_{b,u}  + \frac{1}{2\rho_b}\mathbf{F}_{{\rm HB},b,u}\right) \right)\nonumber \\
    \overset{\left(b\right)}{=} & \mathrm{Tr}\left(  \left(\mathbf{\Sigma}_{b,u} + \frac{1}{2\rho_b}\mathbf{I}_{N_t} + \lambda_b \mathbf{I}_{N_t} \right)^{-2} \mathbf{U}_{b,u}^{\rm H}\right. \nonumber \\
    & \left. \left(\mathbf{M}_{b,u}  + \frac{1}{2\rho_b}\mathbf{F}_{{\rm HB},b,u}\right) \left(\mathbf{M}_{b,u}  + \frac{1}{2\rho_b}\mathbf{F}_{{\rm HB},b,u}\right)^{\rm H}\mathbf{U}_{b,u}  \right)\nonumber \\
    \overset{\left(c\right)}{=} & \mathrm{Tr}\left(  \left(\mathbf{\Sigma}_{b,u} + \frac{1}{2\rho_b}\mathbf{I}_{N_t} + \lambda_b \mathbf{I}_{N_t} \right)^{-2} \mathbf{P}_{b,u} \right),
	\end{align}
    where $\left( a \right)$ holds by calculating the singular value decomposition (SVD) as $\mathbf{Q}_{b,u} = \mathbf{U}_{b,u}\mathbf{\Sigma}_{b,u}\mathbf{U}_{b,u}^{\rm H}$, $\left( b \right)$ is due to the fact that $\mathbf{U}_{b,u}$ is a unitary matrix, which has the property of $ \mathbf{U}_{b,u}^{-1} = \mathbf{U}_{b,u}^{\rm H} $, and $\left( c \right)$ holds by defining $\mathbf{P}_{b,u} = \mathbf{U}_{b,u}^{\rm H} \left(\mathbf{M}_{b,u}  + \frac{1}{2\rho_b}\mathbf{F}_{{\rm HB},b,u}\right) \left(\mathbf{M}_{b,u}  + \frac{1}{2\rho_b}\mathbf{F}_{{\rm HB},b,u}\right)^{\rm H}\mathbf{U}_{b,u}$.
    For the $b$-th AP, the transmit power is expressed as
    \begin{align}\label{Pt_b}
    &\sum_{u=1}^U\left( \mathrm{Tr}\left( \left( \mathbf{F}_{b,u}^{\star}\left( \lambda_b \right) \right)^{\rm H}  \mathbf{F}_{b,u}^{\star}\left( \lambda_b \right) \right) \right) \nonumber \\
    = & \sum_{u=1}^U\left( \mathrm{Tr}\left(  \left(\mathbf{\Sigma}_{b,u} + \frac{1}{2\rho_b}\mathbf{I}_{N_t} + \lambda_b \mathbf{I}_{N_t} \right)^{-2} \mathbf{P}_{b,u} \right) \right) \nonumber \\
    = & \sum_{u=1}^U\sum_{n=1}^{N_t} \frac{\mathbf{P}_{b,u}\left(n,n\right)}{\left(\mathbf{\Sigma}_{b,u}\left(n,n\right) + \frac{1}{2\rho_b} + \lambda_b \right)^2}.
	\end{align}
    We observe that the transmit power is a monotonically decreasing function with respect to the Lagrangian multiplier $\lambda_b$.
    The upper bound on the transmit power of the $b$-th AP can be derived by setting $\lambda_b = 0$, which can be expressed as
    \begin{align}\label{Upb_Pt}
    & \sum_{u=1}^U\left( \mathrm{Tr}\left( \left( \mathbf{F}_{b,u}^{\star}\left( \lambda_b \right) \right)^{\rm H}  \mathbf{F}_{b,u}^{\star}\left( \lambda_b \right) \right) \right) \nonumber \\
    & \leq \sum_{u=1}^U\sum_{n=1}^{N_t} \frac{\mathbf{P}_{b,u}\left(n,n\right)}{\left(\mathbf{\Sigma}_{b,u}\left(n,n\right) + \frac{1}{2\rho_b}\right)^2}.
	\end{align}
    For deriving the optimal $\lambda_b$, we discuss the following two cases:
    \begin{enumerate}[\text{Case} 1: ]
    \item
    If the upper bound $\sum_{u=1}^U\sum_{n=1}^{N_t} \frac{\mathbf{P}_{b,u}\left(n,n\right)}{\left(\mathbf{\Sigma}_{b,u}\left(n,n\right) + \frac{1}{2\rho_b}\right)^2} \leq P_{b,\max}$, the power constraint (\ref{sub_probl_F}b) for the $b$-th AP is naturally satisfied.
    Therefore, we set the optimal $\lambda_b^{\star}$ to $0$.
    \item
    If the upper bound $\sum_{u=1}^U\sum_{n=1}^{N_t} \frac{\mathbf{P}_{b,u}\left(n,n\right)}{\left(\mathbf{\Sigma}_{b,u}\left(n,n\right) + \frac{1}{2\rho_b}\right)^2} > P_{b,\max}$, the bisection search method is executed for determining the optimal $\lambda_b$.
    Specifically, $\lambda_b^{\star}$ is searched within the range $\lambda_b^{\star} \in \left( \lambda_{\rm lb},\lambda_{\rm ub} \right)$, where  $\lambda_{\rm lb}$ is initialized as $0$ and $\lambda_{\rm ub}$ is initialized as $\sqrt{\frac{\sum_{u=1}^U\sum_{n=1}^{N_t} \mathbf{P}_{b,u}\left(n,n\right)}{P_{b,\max}}}$, which is due to $\sum_{u=1}^U\sum_{n=1}^{N_t} \frac{\mathbf{P}_{b,u}\left(n,n\right)}{\left(\mathbf{\Sigma}_{b,u}\left(n,n\right) + \frac{1}{2\rho_b} + \lambda_{\rm ub}\right)^2} < \sum_{u=1}^U\sum_{n=1}^{N_t} \frac{\mathbf{P}_{b,u}\left(n,n\right)}{\lambda_{\rm ub}^2} \overset{\Delta}{=} P_{b,\max}$.
    \end{enumerate}
    Once $\lambda_b^{\star}$ is obtained, the optimal $\mathbf{F}_{b,u}^{\star}\left( \lambda_b^{\star} \right)$ is expressed as
    \begin{align}\label{F_buopt_lbd}
    &\mathbf{F}_{b,u}^{\star}\left( \lambda_b^{\star} \right)\nonumber \\
    &= \left(\mathbf{Q}_{b,u} + \frac{1}{2\rho_b}\mathbf{I}_{N_t} + \lambda_b^{\star} \mathbf{I}_{N_t} \right)^{-1} \left(\mathbf{M}_{b,u}  + \frac{1}{2\rho_b}\mathbf{F}_{{\rm HB},b,u}\right).
	\end{align}
    The procedure for deriving the optimal $\mathcal{F}$ is summarized in Algorithm 1.
\begin{algorithm}[h]
\caption{Lagrangian multiplier method and PDD method for solving the problem (\ref{sub_probl_F})}
    \begin{algorithmic}[2]
        \REQUIRE ~~\\
        $\mathcal{W}$, $\mathcal{G}$, $\mathcal{F}_{\rm RF}$, $\mathcal{F}_{\rm BB}$, the penalty parameters $\left\{\rho_b = \frac{N_t}{100} | \forall b \in \mathcal{B} \right\}$, initial $\mathcal{F}$, Bisection search accuracy $\epsilon$
        \ENSURE
        \STATE Calculate $\mathbf{Q}_{b,u}$ and $\mathbf{M}_{b,u}$, $\forall b \in \mathcal{B}, u \in \mathcal{U}$ according to Proposition 1.
        \FOR{$b = 1 : B$}
        \STATE Calculate the upper bound on transmit power for the $b$-th AP according to (\ref{Upb_Pt});
        \IF {Case 1 is satisfied}
        \STATE Set $\lambda_b^{\star}=0$;
        \ELSE
        \STATE Initialize $\lambda_{\rm ub}$ and $\lambda_{\rm lb}$;
        \STATE Calculate $\lambda_b = \frac{\lambda_{\rm ub} + \lambda_{\rm lb}}{2}$;
        \STATE If $\sum_{u=1}^U\left( \mathrm{Tr}\left( \left( \mathbf{F}_{b,u}^{\star}\left( \lambda_b \right) \right)^{\rm H}  \mathbf{F}_{b,u}^{\star}\left( \lambda_b \right) \right) \right) \geq P_{b,\max}$, set $\lambda_{\rm lb} = \lambda_b$. Otherwise, set $\lambda_{\rm ub} = \lambda_b$;
        \STATE If $\left| \lambda_{\rm ub} - \lambda_{\rm lb} \right| < \epsilon$, set $\lambda_b^{\star} = \lambda_b$ and go to step 12. Otherwise, go back to step 8;
        \ENDIF
        \STATE Calculate $\mathbf{F}_{b,u}^{\star}\left( \lambda_b^{\star} \right)$ according to (\ref{F_buopt_lbd}) $\forall u \in \mathcal{U}$.
        \ENDFOR
        \STATE Calculate $\mathbf{F}_{u} = \left[ \mathbf{F}_{1,u}^{\rm H}, \mathbf{F}_{2,u}^{\rm H},\cdots,\mathbf{F}_{B,u}^{\rm H} \right]^{\rm H}$.
        \label{code:recentEnd}
    \end{algorithmic}
\end{algorithm}
    \subsection{Designing the Analog and Digital Precoding Matrices $\mathcal{F}_{\rm RF}$ and $\mathcal{F}_{\rm BB}$}\label{S3.4}
    We observe that only the penalty terms $\sum_{b=1}^B\sum_{u=1}^U\frac{1}{2\rho_b}\left\| \mathbf{F}_{b,u} - \mathbf{F}_{{\rm HB},b,u} \right\|_F^2$ of the OF (\ref{PDD}a) contain the analog TPC matrices $\mathcal{F}_{\rm RF}$ and the digital TPC matrices $\mathcal{F}_{\rm BB}$.
    Moreover, we have obtained the solution of the auxiliary TPC matrices $\mathcal{F}$.
    Therefore, given that the other variables are fixed, we can derive the closed-form solution of $\mathcal{F}_{\rm RF}$ and $\mathcal{F}_{\rm BB}$.

    For deriving $\mathcal{F}_{{\rm BB}}$, we divide the problem (\ref{PDD}) into $BU$ subproblems.
    For the digital TPC matrix of the $u$-th user at the $b$-th AP, the unconstrained subproblem is formulated by
    \begin{align}\label{sub_probl_F_BB}
    \min_{\mathbf{F}_{{\rm BB},b,u}} \  \left\| \mathbf{F}_{b,u} - \mathbf{F}_{{\rm RF},b}\mathbf{F}_{{\rm BB},b,u} \right\|_F^2.
    \end{align}
    The optimal closed-form solution can be obtained by adopting the well-known least square (LS) method.
    Hence we have
    \begin{align}\label{F_BB_opt}
    \mathbf{F}_{{\rm BB},b,u}^{\star} = \mathbf{F}_{{\rm RF},b}^{\dagger} \mathbf{F}_{b,u}.
    \end{align}

    For deriving $\mathcal{F}_{{\rm RF}}$, we formulate the $B$ subproblems, each of which is expressed as
    \begin{subequations}\label{sub_probl_F_RF}
    \begin{align}
    & \min_{\mathbf{F}_{{\rm RF},b}} \  \left\| \overline{\mathbf{F}}_{b} - \mathbf{F}_{{\rm RF},b}\overline{\mathbf{F}}_{{\rm BB},b} \right\|_F^2 \\
    & \ s.t. \quad \left| \mathbf{F}_{{\rm RF},b}\left( i,j \right) \right| = 1, \forall i \in \mathcal{N}_t, \forall j \in \mathcal{N}_{\rm RF},
    \end{align}
    \end{subequations}
    where $\overline{\mathbf{F}}_{b} = \left[ \mathbf{F}_{b,1},\mathbf{F}_{b,2},\cdots, \mathbf{F}_{b,U} \right]$ and $\overline{\mathbf{F}}_{{\rm BB},b} = \left[ \mathbf{F}_{{\rm BB},b,1},\mathbf{F}_{{\rm BB},b,2},\cdots, \mathbf{F}_{{\rm BB},b,U} \right]$.
    We derive the optimal closed-form update for each subproblem and we update each element of $\mathbf{F}_{{\rm RF},b}$ relying on the classical LS method as \cite{7579557}
    \begin{align}\label{F_RF_opt}
    \mathbf{F}_{{\rm RF},b}^{\star}\left( i,j \right) = e^{j\xi_{i,j}} \mathbf{F}_{\mathrm{RF},b}\left(i,j\right),
    \end{align}
    where $\xi_{i,j} = \angle \left\{ \left(  \mathbf{F}_{\mathrm{RF},b}\left(i,j\right) \overline{\mathbf{F}}_{\mathrm{BB},b}\left(j,:\right) \right)^{\ast} \left(\widetilde{\mathbf{F}}_{ b }\left(i,:\right)\right)^{\text{T}} \right\}$ with $\angle \left\{ \cdot \right\}$ denotes the function which extracts the phase, $\widetilde{\mathbf{F}}_{ b } = \overline{\mathbf{F}}_{b} - \mathbf{F}_{{\rm RF},b}^{-j,\mathrm{col}}\overline{\mathbf{F}}_{{\rm BB},b}^{-j,\mathrm{row}}$ with $\mathbf{F}_{{\rm RF},b}^{-j,\mathrm{col}}$  denoting the submatrix of $\mathbf{F}_{\mathrm{RF},b}$ with the $j$-th column removed and $\overline{\mathbf{F}}_{{\rm BB},b}^{-j,\mathrm{row}}$ denoting the submatrix of $\mathbf{F}_{\mathrm{BB},b}$ with the $j$-th row removed.

    Typically, infinite-resolution phase shifters with fully-connected structure are considered in this paper.
    Note that the LS method can be readily extended to other sub-connected structures \cite{9110865}.
    For instance, the LS method of designing analog TPCs can be easily extended to structures with other type of connections by updating $\mathbf{F}_{{\rm RF},b}^{\star}\left( i,j \right)$ for $\left\{ \left(i,j\right) | \forall \left( i,j \right) \in \mathcal{D} \right\}$, where $ \mathcal{D} $ denotes the specific connection set of phase shifters, namely those that are connected to RF chains and antennas.
    For finite-resolution phase shifters, we can extend the LS method by quantizing $\mathbf{F}_{{\rm RF},b}^{\star}\left( i,j \right) = \mathcal{Q} \left[ e^{j\xi_{i,j}} \mathbf{F}_{\mathrm{RF},b}\left(i,j\right) \right]$ during the update of analog TPCs, where $\mathcal{Q} \left( \cdot \right)$ stands for the quantization function.
    \subsection{Overall Algorithm and Convergence Analysis}\label{S3.5}
    In this subsection, we will describe the overall algorithm.
    Since our proposed method relies on the BCD algorithm, an iterative update procedure is utilized.
    Specifically, we iteratively calculate $\mathcal{G}$, $\mathcal{W}$, $\mathcal{F}$, $\mathcal{F}_{\rm BB}$ and $\mathcal{F}_{\rm RF}$ until convergence is realized on basis of the aforementioned derivation.
    Now we present the initialization of the algorithm.
    The set $\mathcal{S}$ derived in the $n$-th iteration is denoted as $\mathcal{S}^{\left( n \right)}$.
    Then the initialization of $\mathcal{F}^{\left( 0 \right)}$, $\mathcal{F}_{\rm BB}^{\left( 0 \right)}$ and $\mathcal{F}_{\rm RF}^{\left( 0 \right)}$ is discussed as follows.

    $\mathcal{F}^{\left( 0 \right)}$: We first calculate the SVD of all channels between the $b$-th AP and the $u$-th user as $\mathbf{H}_{b,u} = \widetilde{\mathbf{U}}_{b,u}\widetilde{\mathbf{\Sigma}}_{b,u}\widetilde{\mathbf{V}}_{b,u}^{\rm H}$.
    Then $\mathbf{F}_u^{\left( 0 \right)}$ of the $u$-th user in $\mathcal{F}^{\left( 0 \right)}$ is set to $\mathbf{F}_u^{\left( 0 \right)} = \left[ \left(\mathbf{F}_{1,u}^{\left( 0 \right)}\right)^{\rm H},\left(\mathbf{F}_{2,u}^{\left( 0 \right)}\right)^{\rm H},\cdots,\right. $ $ \left.\left(\mathbf{F}_{B,u}^{\left( 0 \right)}\right)^{\rm H} \right]^{\rm H}$, where $\mathbf{F}_{b,u}^{\left( 0 \right)} = \frac{\sqrt{P_{b,\max}}\widetilde{\mathbf{V}}_{b,u}\left( :, 1:N_s \right)}{\sqrt{\sum_{u=1}^U\left\|\widetilde{\mathbf{V}}_{b,u}\left( :, 1:N_s \right)\right\|_F^2}}$ satisfies the power constraint (\ref{sub_probl_F}b).

    $\mathcal{F}_{\rm RF}^{\left( 0 \right)}$: each element in $\mathbf{F}_{{\rm RF},b}^{\left( 0 \right)}$ is set to $e^{j\phi_{b,i,j}}$ with $\phi_{b,i,j}$ denoting random phases chosen from $\left[-\pi, \pi \right]$, $\forall i \in \mathcal{N}_t, \forall j \in \mathcal{N}_{\rm RF}, \forall b \in \mathcal{B}$.

    $\mathcal{F}_{\rm BB}^{\left( 0 \right)}$: recalling (\ref{F_BB_opt}), the initial $\mathcal{F}_{{\rm BB},b,u}^{\left( 0 \right)}$ is set to $\mathcal{F}_{{\rm BB},b,u}^{\left( 0 \right)} = \frac{\sqrt{P_{b,\max}}\left(\mathbf{F}_{{\rm RF},b}^{\left( 0 \right)}\right)^{\dagger} \mathbf{F}_{b,u}^{\left( 0 \right)}}{\sqrt{\sum_{u=1}^U \left\| \mathbf{F}_{{\rm RF},b}^{\left( 0 \right)}\left(\mathbf{F}_{{\rm RF},b}^{\left( 0 \right)}\right)^{\dagger} \mathbf{F}_{b,u}^{\left( 0 \right)}\right\|_F^2}}$, which satisfies the power constraint for each AP.

    Note that when the convergence condition is met, $\mathcal{F}_{\rm BB}$ may not satisfy the power constraint for each AP.
    Therefore, we derive the final $\mathbf{F}_{{\rm BB},b,u}^{\star}$, $\forall b \in \mathcal{B}, u \in \mathcal{U}$, as $\mathbf{F}_{{\rm BB},b,u}^{\star} = \frac{\sqrt{P_{b,\max}}\mathbf{F}_{{\rm BB},b,u}^{\star}}{\sqrt{\sum_{u=1}^U \left\| \mathbf{F}_{{\rm RF},b}^{\star}\mathbf{F}_{{\rm BB},b,u}^{\star} \right\|_F^2}}$ \cite{8310586,8331836,7579557}.
    The overall procedure is summarized in Algorithm 2, where $\tilde{R}$ denotes the OF (11a).
\begin{algorithm}[h]
\caption{Overall algorithm for hybrid precoding design in cell-free MIMO}
    \begin{algorithmic}[2]
        \REQUIRE ~~\\
        Initial $\mathcal{F}_{\rm RF}^{\left( 0 \right)}$, $\mathcal{F}_{\rm BB}^{\left( 0 \right)}$, $\mathcal{F}^{\left( 0 \right)}$, iteration index $n=0$, convergency accuracy $\xi$;
        \ENSURE
        \WHILE{$ \left| \tilde{R}^{\left( n \right)} - \tilde{R}^{\left( n - 1 \right)} \right| \geq \xi$}
        \STATE $n = n + 1$;
        \STATE Calculate $\mathcal{G}^{\left( n \right)}$ according to (\ref{G_uopt}) with other variables fixed;
        \STATE Calculate $\mathcal{W}^{\left( n \right)}$ according to (\ref{W_uopt}) with other variables fixed;
        \STATE Calculate $\mathcal{F}^{\left( n \right)} $ according to Algorithm 1 with other variables fixed;
        \STATE Calculate $\mathcal{F}_{\rm BB}^{\left( n \right)} $ according to (\ref{F_BB_opt}) with other variables fixed;
        \STATE Calculate $\mathcal{F}_{\rm RF}^{\left( n \right)}$ according to (\ref{F_RF_opt}) with other variables fixed;
        \ENDWHILE
        \STATE If $\sum_{u=1}^U\left\|\mathbf{F}_{{\rm RF},b}^{\star}\mathbf{F}_{{\rm BB},b,u}^{\star}\right\|_F^2 > P_{b,\max}$, calculate $\mathbf{F}_{{\rm BB},b,u}^{\star}$ as $\mathbf{F}_{{\rm BB},b,u}^{\star} = \frac{\sqrt{P_{b,\max}}\mathbf{F}_{{\rm BB},b,u}^{\star}}{\sqrt{\sum_{u=1}^U\left\|\mathbf{F}_{{\rm RF},b}^{\star}\mathbf{F}_{{\rm BB},b,u}^{\star}\right\|_F^2}}$, $\forall u \in \mathcal{U}$, $\forall b \in \mathcal{B}$.
        \label{code:recentEnd}
    \end{algorithmic}
\end{algorithm}

    Next, we will demonstrate the convergence of our proposed algorithm.
    According to the characteristics of the BCD algorithm, when we derive one set of variables with the others fixed, the OF (\ref{PDD}a) increases or at least maintains its value.
    Hence, the OF (\ref{PDD}a) is a non-decreasing function in our design.

    In order to demonstrate the convergence of the proposed method, we still have to prove that the OF (\ref{PDD}a) is upper-bounded.
    We consider an ideal unconstrained cell-free MIMO scenario, where there is no IUI.
    The TPC matrices of this scenario can be directly obtained by the singular value decomposition (SVD) for each user at each AP.
    Therefore, (\ref{PDD}a) can indeed achieve convergence.
    The proposed algorithm converges to at least locally optimal solutions of the transformed problem (\ref{PDD}).
    \subsection{Complexity Analysis}\label{S3.6}
    During the design of $\mathcal{F}$, the dominant complexity contribution is that of calculating the inverse matrix $\left(\mathbf{Q}_{b,u} + \frac{1}{2\rho_b}\mathbf{I}_{N_t} + \lambda_b^{\star} \mathbf{I}_{N_t} \right)^{-1}$, $\forall b \in \mathcal{B}, \forall u \in \mathcal{U}$, which has the complexity order of $\mathcal{O}\left( N_t^3 \right)$.
    The bisection search of $\lambda_b^{\star}$ for the $b$-th AP has to repeat the calculation $T_b = \log_2\left( \frac{\lambda_{\rm ub} - \lambda_{\rm lb}}{\epsilon} \right)$ times.
    Therefore, the complexity order of optimizing $\mathcal{F}$ is $\mathcal{O}\left( BUTN_t^3 \right)$, where $T = \sum_{b=1}^B T_b$.
    For designing $\mathcal{F}_{\rm RF}$ and $\mathcal{F}_{\rm BB}$, the complexity order is $\mathcal{O}\left[ B\left( N_tN_{\rm RF}UN_s \right. \right. $ $\left. \left. + N_{\rm RF}^3 \right) \right]$.
    If the number of iterations required for convergence is $N_{\max}$, the total complexity order is given by $\mathcal{O}\left[ N_{\max}\left( BUTN_t^3 + B\left( N_tN_{\rm RF}UN_s + N_{\rm RF}^3 \right) \right) \right]$.

    Note that our proposed algorithm can be modified for the cell-free MIMO DL, where fully digital TPC structures are adopted.
    In the next section, we will design the fully digital TPC matrices by modifying our proposed algorithm.
    \section{Modified Algorithm for Fully Digital Precoding in cell-free MIMO}\label{S4}
    In this section, we consider a scenario where each AP in the cell-free MIMO DL is equipped with fully digital TPC.
    The fully digital TPC matrices are designed by modifying the algorithm proposed in Section \ref{S3}.
    \subsection{Problem Formulation}\label{S4.1}
    The set of fully digital TPC matrices is defined by $\mathcal{F}_{\rm FD} = \left\{ \mathbf{F}_{{\rm FD},u} | \forall u \in \mathcal{U} \right\}$ and $\mathbf{F}_{{\rm FD},u} = \left[ \mathbf{F}_{{\rm FD},1,u}^{\rm H}, \mathbf{F}_{{\rm FD},2,u}^{\rm H}, \cdots, \mathbf{F}_{{\rm FD},B,u}^{\rm H} \right]^{\rm H}$.
    The WSR maximization problem is formulated by
    \begin{subequations}\label{ObjSR_FD_v1}
    \begin{align}
    \max_{ \mathcal{F}_{\rm FD} } \quad &  R_{\rm FD} \\
    s.t. \quad \ &  \sum_{u=1}^U \mathrm{Tr}\left( \mathbf{F}_{{\rm FD},u}\mathbf{F}_{{\rm FD},u}^{\rm H} \right) \leq P_{b,\max}, \forall b \in \mathcal{B},
    \end{align}
    \end{subequations}
    where $R_{\rm FD} = \sum_{u=1}^{U}\omega_u \log \left| \mathbf{I}_{N_r} + \mathbf{J}^{-1}_{{\rm FD},u}\mathbf{H}_{u}\mathbf{F}_{{\rm FD},u}\mathbf{F}_{{\rm FD},u}^{\mathrm{H}}\mathbf{H}_{u}^{\mathrm{H}} \right|$ and $\mathbf{J}_{{\rm FD},u} = \sum_{j=1,j\neq u}^U\mathbf{H}_{u}\mathbf{F}_{{\rm FD},j}\mathbf{F}_{{\rm FD},j}^{\rm H}$ $\mathbf{H}_{u}^{\rm H} + \sigma^2\mathbf{I}_{N_r}$.
    \subsection{Modified Precoding Algorithm}\label{S4.2}
Similar to the problem transformation from Section III-A to Section III-C, we can obtain the unconstrained problem by exploiting the equivalent WMMSE transformation and BCD algorithm for the fully digital scenario.
Specifically, we iteratively update the introduced linear combiners $\left\{ \mathbf{G}_u | \forall u \in \mathcal{U}\right\}$ in $\mathcal{G}$, the introduced weighting matrices $\left\{ \mathbf{W}_u | \forall u \in \mathcal{U} \right\}$ in $\mathcal{W}$ and the fully digital TPC matrices $\left\{ \mathbf{F}_{{\rm FD},u} | \forall u \in \mathcal{U} \right\}$ in $\mathcal{F}_{\rm FD}$.
During the update, $\mathbf{G}_{u}$ and $\mathbf{W}_{u}$ can be calculated by exploiting the partial derivative based method, which is similar to (13) and (14) in the hybrid TPC design.
Then we propose to design fully digital TPC matrices by exploiting the Lagrangian multiplier method.
    The transformed unconstrained problem is expressed by
    \begin{align}\label{sub_probl_FFD_Lagrangian_v1}
    \min_{\left\{ \mu_b | \forall b \in \mathcal{B} \right\},\mathcal{F}_{\rm FD}} \ & \sum_{u=1}^U\omega_u\left[\mathrm{Tr}\left( \mathbf{F}_{{\rm FD},u}^{\rm H} \mathbf{A}_u \mathbf{F}_{{\rm FD},u} \right)\right] \nonumber\\
    & - \sum_{u=1}^U\omega_u\left[\mathrm{Tr}\left( \mathbf{F}_{{\rm FD},u}^{\rm H}\mathbf{H}_{u}^{\rm H}\mathbf{G}_{u}\mathbf{W}_{u} \right) \right] \nonumber\\
    & - \sum_{u=1}^U\omega_u\left[\mathrm{Tr}\left( \mathbf{W}_{u}\mathbf{G}_{u}^{\rm H}\mathbf{H}_{u}\mathbf{F}_{{\rm FD},u} \right)\right] \nonumber\\
    & + \sum_{b=1}^B \mu_b \left[ \sum_{u=1}^U \mathrm{Tr}\left( \mathbf{F}_{{\rm FD},b,u}\mathbf{F}_{{\rm FD},b,u}^{\rm H}\right) - P_{b,\max} \right],
    \end{align}
    where $\left\{ \mu_b \geq 0 | \forall b \in \mathcal{B} \right\}$ is the set of Lagrangian multipliers introduced.
    Since no inter-user interference exists in any of the UC cluster, our problem formulation is different from that of the existing literature \cite{5756489,alexandropoulos2013reconfigurable,7509379}.
    Similar to Proposition 1, stated for deriving the optimal $\mathcal{F}_{\rm FD}$, we have the following Proposition 2.

    \textbf{Proposition 2}:
    We propose to iteratively update $\left\{ \mathbf{F}_{\mathrm{FD},b,u} | \forall b \in \mathcal{B}, \forall u \in \mathcal{U} \right\}$ in each iteration of the BCD algorithm.
    The problem (\ref{sub_probl_FFD_Lagrangian_v1}) can be rewritten as
\begin{align}\label{sub_probl_FFD_Lagrangian_v2}
    & \min_{\left\{ \mathbf{F}_{\mathrm{FD},b,u} | \forall b \in \mathcal{B}, \forall u \in \mathcal{U} \right\}} \nonumber \\
    & \sum_{u=1}^U\left[\mathrm{Tr}\left( \mathbf{F}_{{\rm FD},b,u}^{\rm H} \left(\mathbf{Q}_{b,u} + \mu_b \mathbf{I}_{N_t} \right) \mathbf{F}_{{\rm FD}b,u} \right)\right] \nonumber\\
    & + \sum_{i=1,i\neq b}^B\sum_{u=1}^U\left[\mathrm{Tr}\left( \mu_b \mathbf{F}_{{\rm FD},i,u}^{\rm H} \mathbf{F}_{{\rm FD}i,u} \right)\right] \nonumber \\
    & - \sum_{u=1}^U\left[\mathrm{Tr}\left(\mathbf{M}_{b,u}^{\rm H}\mathbf{F}_{b,u} \right)\right] - \sum_{u=1}^U\left[\mathrm{Tr}\left( \mathbf{F}_{b,u}^{\rm H}\mathbf{M}_{b,u} \right) \right]\nonumber \\
    & - \sum_{i=1,i\neq b}^B\sum_{u=1}^U\left[\mathrm{Tr}\left(\mathbf{C}_{i,u}^{\rm H}\mathbf{F}_{i,u} \right)\right] - \sum_{i=1,i\neq b}^B\sum_{u=1}^U\left[\mathrm{Tr}\left( \mathbf{F}_{i,u}^{\rm H}\mathbf{C}_{i,u} \right) \right].
    \end{align}
    \begin{IEEEproof}
        The proof is similar to that in Appendix A, except that the penalty terms are eliminated.
    \end{IEEEproof}
    Based on Proposition 2, the optimal closed-form $\mathbf{F}_{\mathrm{FD},b,u}\left( \mu_b \right)$ can be formulated as
    \begin{align}\label{F_FD_buopt}
    \mathbf{F}_{{\rm FD},b,u}^{\star}\left( \mu_b \right) = \left(\mathbf{Q}_{b,u} + \mu_b \mathbf{I}_{N_t} \right)^{\dagger} \mathbf{M}_{b,u}.
	\end{align}
    The determination of the optimal Lagrangian multipliers $\left\{ \mu_b | \forall b \in \mathcal{B} \right\}$ is different from that of the hybrid TPC scenario due to the fact that $\mathbf{Q}_{b,u}$ may be a low-rank matrix.
    For obtaining the optimal Lagrangian multipliers, we have the following Lemma concerning the rank of $\mathbf{Q}_{b,u}$.

    \textbf{Lemma 1}: $\text{rank}\left(\mathbf{Q}_{b,u}\right) =  N_s$.
    \begin{IEEEproof}
    Please see Appendix B.
    \end{IEEEproof}
    For $N_s = N_t$, the matrix $\mathbf{Q}_{b,u}$ is of full rank, thus the derivation for $\left\{ \mu_b | \forall b \in \mathcal{B} \right\}$ is the same as that of the hybrid TPC design.
    Considering $N_s < N_t$, on the basis of Lemma 1, the transmit power is modified according to
    \begin{align}\label{Upb_Pt2_1}
    & \sum_{u=1}^U\left( \mathrm{Tr}\left( \left( \mathbf{F}_{{\rm FD},b,u}^{\star}\left( \mu_b \right) \right)^{\rm H}  \mathbf{F}_{{\rm FD},b,u}^{\star}\left( \mu_b \right) \right) \right)  \nonumber\\
    &= \left\{
    \begin{aligned}
     &  \sum_{u=1}^U\sum_{n=1}^{N_s} \frac{\mathbf{P}_{{\rm FD},b,u}\left(n,n\right)}{\left(\mathbf{\Sigma}_{b,u}\left(n,n\right) + \mu_b \right)^2} \\
     & + \sum_{u=1}^U\sum_{n=N_s+1}^{N_t} \frac{\mathbf{P}_{{\rm FD},b,u}\left(n,n\right)}{\mu_b^2} ,\quad\quad\quad\quad \text{for } \mu_b \neq 0,\\
     &  \sum_{u=1}^U\sum_{n=1}^{N_s} \frac{\mathbf{P}_{{\rm FD},b,u}\left(n,n\right)}{\left(\mathbf{\Sigma}_{b,u}\left(n,n\right) \right)^2}, \ \ \quad\quad\quad\quad\quad\quad \text{for } \mu_b = 0.
    \end{aligned}
    \right.
	\end{align}
    In contrast to the hybrid TPC algorithm, the transmit power is not upper bounded by $\sum_{u=1}^U$ $\left( \mathrm{Tr}\left( \left( \mathbf{F}_{{\rm FD},b,u}^{\star}\left( 0 \right) \right)^{\rm H}  \mathbf{F}_{{\rm FD},b,u}^{\star}\left( 0 \right) \right) \right)$.
    Therefore, we adopt the bisection search for deriving the optimal Lagrangian multipliers  $\left\{ \mu_b^{\star} | \forall b \in \mathcal{B} \right\}$.
    According to (\ref{Upb_Pt2_1}), since the transmit power $\sum_{u=1}^U$ $\left( \mathrm{Tr}\left( \left( \mathbf{F}_{{\rm FD},b,u}^{\star}\left( \mu_b \right) \right)^{\rm H}  \mathbf{F}_{{\rm FD},b,u}^{\star}\left( \mu_b \right) \right) \right)$ is a monotonically decreasing function for $\mu_b > 0$, the optimal Lagrangian multiplier $\mu_b^{\star}$ is sought within the range $\mu_b^{\star} \in \left( \mu_{\rm lb},\mu_{\rm ub} \right)$ through the bisection search.
    During the search, $\mu_{\rm lb}$ is initialized to a small positive value and $\mu_{\rm ub}$ is initialized as $\sqrt{\frac{\sum_{u=1}^U\sum_{n=1}^{N_t} \mathbf{P}_{{\rm FD},b,u}\left(n,n\right)}{P_{b,\max}}}$, which is due to $\sum_{u=1}^U\sum_{n=1}^{N_t} \frac{\mathbf{P}_{{\rm FD},b,u}\left(n,n\right)}{\left(\mathbf{\Sigma}_{b,u}\left(n,n\right) + \mu_{\rm ub}\right)^2} < \sum_{u=1}^U\sum_{n=1}^{N_t} \frac{\mathbf{P}_{{\rm FD},b,u}\left(n,n\right)}{\mu_{\rm ub}^2} \overset{\Delta}{=} P_{b,\max}$.
    Then the optimal fully digital TPC matrix for the $u$-th user at the $b$-th AP $\mathbf{F}_{b,u}^{\star}\left( \mu_b^{\star} \right)$ is expressed as
    \begin{align}\label{FFD_buopt_lbd}
    \mathbf{F}_{{\rm FD},b,u}^{\star}\left( \mu_b^{\star} \right) = \left(\mathbf{Q}_{b,u} + \mu_b^{\star} \mathbf{I}_{N_t} \right)^{\dagger} \mathbf{M}_{b,u}.
	\end{align}
    According to Section \ref{S3.5}, the convergence of the fully digital TPC design can also be proved.
    In the next section, we will demonstrate the superiority of our proposed algorithms through simulation results.

    \section{Numerical Results}\label{S5}
    In our simulations, unless otherwise stated, there are $B = 2$ APs and $U=4$ users in the cell-free MIMO DL considered.
    Each AP has $N_t = 8\times8$ TAs and each user has $N_r=2\times2$ RAs.
    The number of RF chains at each AP is set to $N_{\rm RF} = 8$.
    The number of data streams for each user is set to $N_s = 2$.
    The weighting coefficients for the users are set to $\omega_1 = \cdots = \omega_U = 1$.
    The maximum transmit power for the APs is $P_{1,\max} = \cdots = P_{B,\max} = P_{\max} = 30$dBm.
    The penalty parameters introduced are set to $\rho_1 = \cdots = \rho_B = \frac{100}{N_t}$ \cite{9110865}.
    The thresholds for Algorithm 1 and Algorithm 2 are set to $\epsilon = 10^{-6}$ and $\xi = 10^{-4}$.
    The narrow-band mmWave channels have $L=8$ paths.
    The azimuth AoA $\phi^{\mathrm{r}}_{b,u}$ and AoD $\phi^{\mathrm{t}}_{b,u}$ are uniformly distributed in the interval $\left( -\pi,\pi \right]$, while the elevation AoA $\theta^{\mathrm{r}}_{b,u}$ and AoD $\theta^{\mathrm{t}}_{b,u}$ are uniformly distributed in the interval $\left( -\frac{\pi}{2},\frac{\pi}{2} \right]$ \cite{TWC_AAhmed_LimitedHybridPrecoding}.
    The noise power is set to $\sigma^2 = 1$.

 	\begin{figure}[t]
		\center{\includegraphics[width=0.5\textwidth]{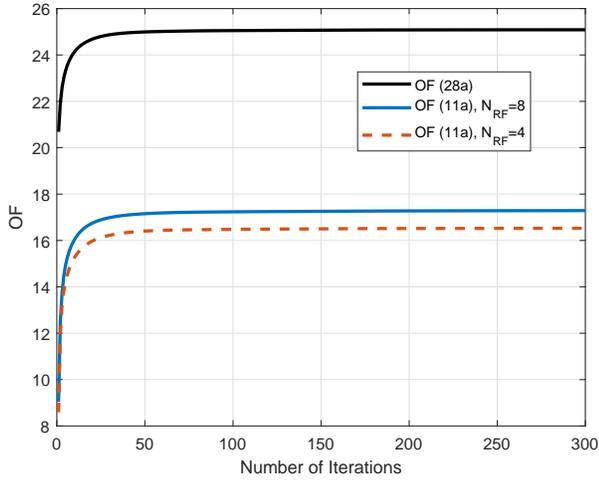}}
		\caption{WSR v.s. the number of iterations.}
		\label{SR_Conv}
	\end{figure}
    In Fig. \ref{SR_Conv}, we evaluate the validity of our convergence analysis for both the hybrid and for the fully digital TPC design.
    It is shown that $N_{\max} = 100$ iterations are enough for achieving near-perfect convergence for both the hybrid and for the fully digital TPC.
    This observation confirms the low complexity of the proposed algorithm.
 	\begin{figure}[t]
		\center{\includegraphics[width=0.5\textwidth]{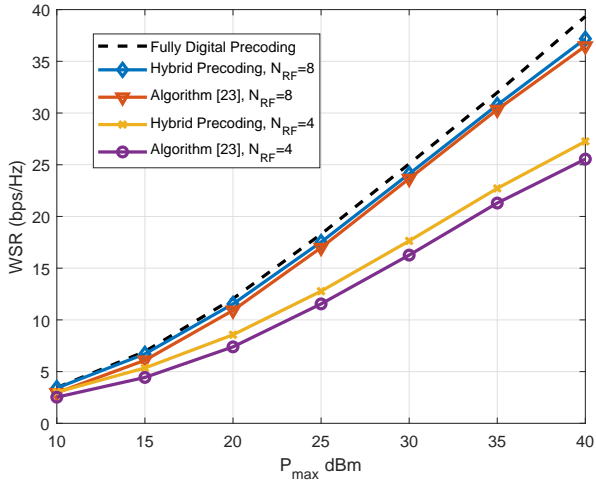}}
		\caption{WSR v.s. transmit power at each AP.}
		\label{SR_Pm}
	\end{figure}

    In Fig. \ref{SR_Pm}, we characterize both the fully digital and the hybrid TPC matrix design in the cell-free MIMO DL vs. the maximum transmit power.
    Observe in Fig. \ref{SR_Pm} that the system's WSR is increased, when the transmit power increases from 10dBm to 40dBm.
    We also observe that as the the number of RF chains is increased, the WSR also increases.
    When the number of RF chains is $N_{\rm RF} = 8$, the cell-free MIMO relying on hybrid TPC achieves a WSR very close to that of fully digital TPC, which verifies the benefits of hybrid TPCs in the cell-free MIMO DL.
    For verifying the superiority of our proposed algorithm, we compare it to the algorithm of \cite{8676377}, where the hybrid TPC matrices are obtained through decomposing the fully digital TPC matrices derived into hybrid ones through subspace decomposition.
    As shown in Fig. \ref{SR_Pm}, the proposed algorithm outperforms the algorithm of \cite{8676377} in terms of the WSR performance.
    Moreover, the algorithm of [23] has higher computational complexity primarily owing to the iterative update of the analog and digital TPC matrices during the hybrid TPC design.

 	\begin{figure}[t]
		\center{\includegraphics[width=0.5\textwidth]{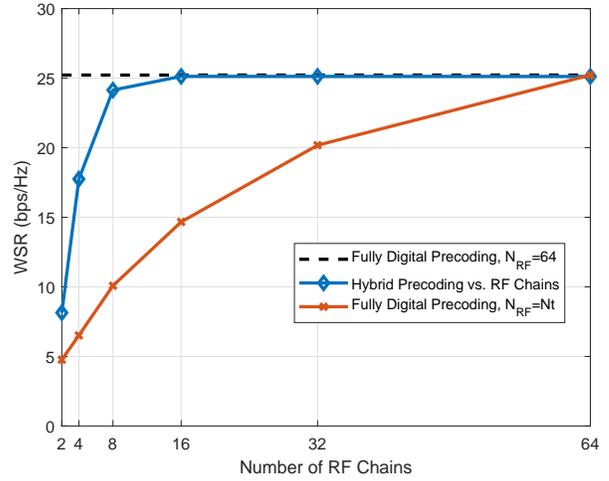}}
		\caption{WSR v.s. the number of RF chains.}
		\label{SR_NRF}
	\end{figure}
    In order to investigate the performance of hybrid TPCs, we increase the number of RF chains from $N_{\rm RF} = 2$ to  $N_{\rm RF} = 64$.
    Observe in Fig. \ref{SR_NRF} that the WSR of hybrid TPCs increases with $N_{\rm RF}$ and a near-optimal WSR is attained for $N_{\rm RF} \geq 16$.
    We also observe that when the number of RF chains in the hybrid TPC is equal to the number of antennas in the fully digital TPC, the WSR of hybrid TPC becomes significantly better when the number of RF chains is limited.
    This further demonstrates the superiority of employing hybrid TPCs in the cell-free MIMO DL.

 	\begin{figure}[t]
		\center{\includegraphics[width=0.5\textwidth]{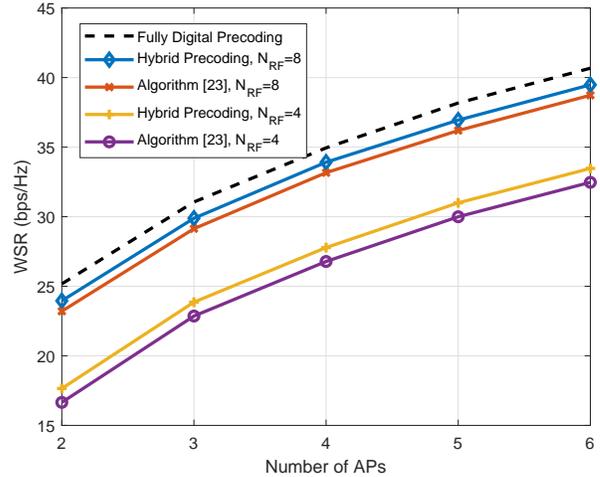}}
		\caption{WSR v.s. the number of APs.}
		\label{SR_B}
	\end{figure}
 	\begin{figure}[t]
		\center{\includegraphics[width=0.5\textwidth]{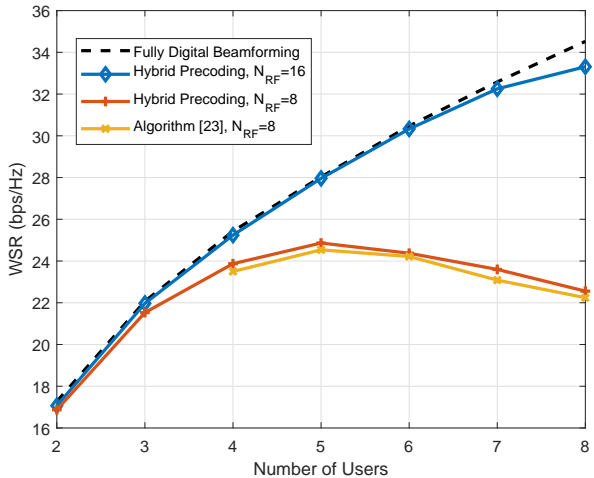}}
		\caption{WSR v.s. the number of users.}
		\label{SR_user}
	\end{figure}

    In Fig. \ref{SR_B}, we investigate the WSR vs. the number of APs in the cell-free MIMO DL.
    In addition to the conclusion drawn from the aforementioned simulations, we observe that the WSR increases as the number of APs increases from $2$ to $6$.
    This means that a large number of APs deployed in the cell-free MIMO DL has the potential of drastically improving the WSR performance.

    Fig. \ref{SR_user} portrays the WSR vs. the number of users.
    Note that the subspace decomposition algorithm cannot be executed when $UN_s < N_{\rm RF}$, hence we do not compare our proposed algorithm to it in those scenarios.
    As illustrated in Fig. \ref{SR_user}, when the number of users increases from $2$ to $8$, the WSR increases first and then decreases for $N_{\rm RF} = 8$.
    This is due to the IUI, which is increased with the number of users, while the number of RF chains is low.

 	\begin{figure}[t]
		\center{\includegraphics[width=0.5\textwidth]{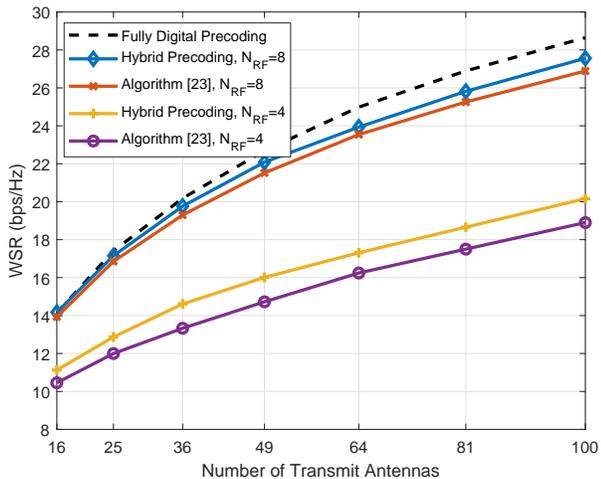}}
		\caption{WSR v.s. the number of TAs.}
		\label{SR_TA}
	\end{figure}
    We then portray the WSR vs. the number of TAs in Fig. \ref{SR_TA}.
    As the number of TAs increases from $N_t = 4 \times 4 = 16$ to $N_t = 10\times 10 =100$, the WSR is nearly doubled.
    This is because more TAs are able to attain a higher TPC gain and antenna diversity for the cell-free MIMO DL.

 	\begin{figure}[t]
		\center{\includegraphics[width=0.5\textwidth]{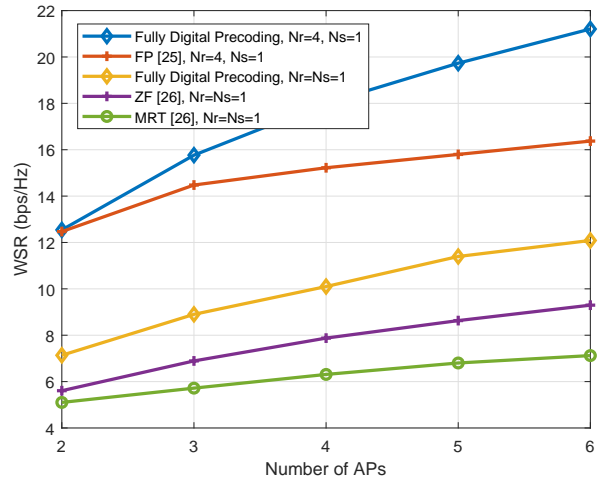}}
		\caption{WSR v.s. the number of APs for the fully digital TPC scenario.}
		\label{Method_Comp}
	\end{figure}
 	\begin{figure}[t]
		\center{\includegraphics[width=0.5\textwidth]{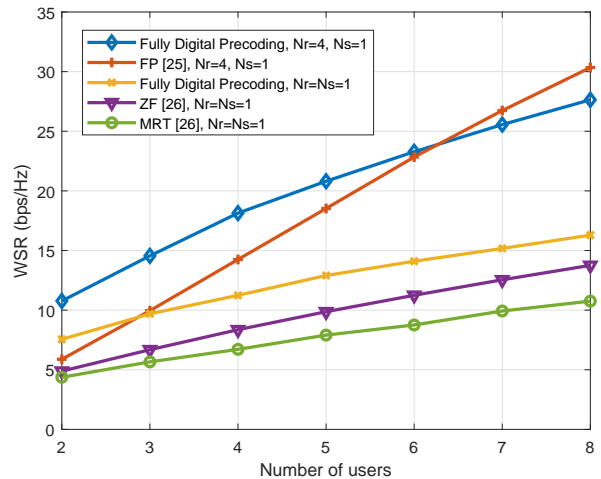}}
		\caption{WSR v.s. the number of users for the fully digital TPC scenario.}
		\label{Method_Comp_U}
	\end{figure}

    As illustrated in Fig. \ref{Method_Comp}, our proposed fully digital TPC shows a better WSR performance compared to its conventional counterparts as the number of APs increases from 2 to 6.
    For explicitly showing the relationships between our proposed algorithm and its counterparts, we set $N_t = 4 \times 4 = 16$ in this figure.
    Furthermore, we set the number of data streams for each user to $N_s = 1$, because the problem transformation in \cite{2002.03744v2} using FP is only suitable for such a case.
    We set $N_r = N_s = 1 $, when we compare our proposed algorithm to the traditional ZF algorithm and to the MRT algorithm \cite{9069486}.
    We observe that as the number of APs increases, the WSR advantage of the proposed algorithm becomes higher.

    We also show the performance comparisons of fully digital TPC vs. the number of users in Fig. \ref{Method_Comp_U}.
    We increase $U$ from $2$ to $8$.
    It is shown in Fig.10 that the FP [25] method with $N_r=4$, $N_s=1$ outperforms the proposed fully digital TPC design when the number of users exceeds 6.
    \section{Summary and Conclusion}\label{S6}
    A powerful hybrid TPC was conceived for the mmWave cell-free MIMO DL.
    A WSR maximization problem was formulated subject to a realistic power constraint for each AP and the constant-modulus constraint for the phase shifters of the analog TPCs.
    Then we transformed it into an equivalent WMMSE problem by exploiting the UC nature of the cell-free MIMO DL.
    We proposed a low complexity BCD algorithm for iteratively solving the transformed problem.
    We then also further extended our proposed algorithm to the fully digital cell-free MIMO scenario.
    Additionally, the convergence and computational complexity analysis of our proposed algorithm were presented.
    Our simulation results demonstrated the superiority of the proposed algorithms both in hybrid and in fully digital TPCs.
    \begin{appendices}
    \section{Proof of Proposition 1}
    In order to complete the proof, we rewrite the terms in (\ref{sub_probl_F_Lagrangian_v2}) as follows.
    The expression in the $\mathrm{Tr}\left( \cdot \right)$ function of the first term in (\ref{sub_probl_F_Lagrangian_v2}) can be  rewritten as
    \begin{align}\label{rewrite_1term}
    \mathbf{F}_{u}^{\rm H} \mathbf{A}_u \mathbf{F}_{u}\nonumber = & \mathbf{F}_{u}^{\rm H} \mathbf{A}_{u}^{\frac{1}{2}}\left(\mathbf{A}_{u}^{\frac{1}{2}}\right)^{\rm H}\mathbf{F}_{u} \nonumber \\
    = & \begin{bmatrix} \mathbf{F}_{1,u}^{\rm H} & \mathbf{F}_{2,u}^{\rm H} & \cdots \mathbf{F}_{B,u}^{\rm H} \end{bmatrix}
    \begin{bmatrix} \widehat{\mathbf{A}}_{1,u} \\ \widehat{\mathbf{A}}_{2,u} \\ \vdots \\ \widehat{\mathbf{A}}_{B,u} \end{bmatrix}\nonumber\\
    & \begin{bmatrix} \widehat{\mathbf{A}}_{1,u}^{\rm H} & \widehat{\mathbf{A}}_{2,u}^{\rm H} & \cdots \widehat{\mathbf{A}}_{B,u}^{\rm H} \end{bmatrix}
    \begin{bmatrix} \mathbf{F}_{1,u}\\ \mathbf{F}_{2,u}^{\rm H} \\ \vdots \\ \mathbf{F}_{B,u} \end{bmatrix} \nonumber \\
    = & \left( \sum_{b=1}^B \mathbf{F}_{b,u}^{\rm H}\widehat{\mathbf{A}}_{b,u} \right) \left(\sum_{b=1}^B \mathbf{F}_{b,u}^{\rm H}\widehat{\mathbf{A}}_{b,u}\right)^{\rm H} \nonumber \\
    = & \mathbf{F}_{b,u}^{\rm H}\widehat{\mathbf{A}}_{b,u} \widehat{\mathbf{A}}_{b,u}^{\rm H} \mathbf{F}_{b,u} + \mathbf{F}_{b,u}^{\rm H}\widehat{\mathbf{A}}_{b,u}\sum_{i=1,i\neq b}^{B}\widehat{\mathbf{A}}_{i,u}^{\rm H}\mathbf{F}_{i,u} \nonumber \\
    & + \sum_{i=1,i\neq b}^{B}\mathbf{F}_{i,u}^{\rm H}\widehat{\mathbf{A}}_{i,u} \widehat{\mathbf{A}}_{b,u}^{\rm H} \mathbf{F}_{b,u} \nonumber\\
    & + \sum_{i=1,i\neq b}^{B}\mathbf{F}_{i,u}^{\rm H}\widehat{\mathbf{A}}_{i,u}\sum_{i=1,i\neq b}^{B}\widehat{\mathbf{A}}_{i,u}^{\rm H}\mathbf{F}_{i,u}.
    \end{align}
    Similar to (\ref{rewrite_1term}), the expression in the $\mathrm{Tr}\left( \cdot \right)$ function of the second term in (\ref{sub_probl_F_Lagrangian_v2}) can be rewritten as
    \begin{align}\label{rewrite_2term}
    \mathbf{F}_{u}^{\rm H}\mathbf{H}_{u}^{\rm H}\mathbf{G}_{u}\mathbf{W}_{u} = \frac{1}{\omega_u}\mathbf{F}_{u}^{\rm H}\mathbf{C}_{u} = \frac{1}{\omega_u} \sum_{b=1}^B \mathbf{F}_{b,u}^{\rm H}\mathbf{C}_{b,u}.
    \end{align}
    The third term in (\ref{sub_probl_F_Lagrangian_v2}) can be readily derived by calculating the conjugate transpose of (\ref{rewrite_2term}).
    Furthermore, we expand the Frobenius norm in the fourth term in (\ref{sub_probl_F_Lagrangian_v2}) as
    \begin{align}\label{rewrite_4term}
    & \left\| \mathbf{F}_{b,u} - \mathbf{F}_{{\rm HB},b,u} \right\|_F^2  \nonumber\\
    = & \mathrm{Tr}\left( \left(\mathbf{F}_{b,u} - \mathbf{F}_{{\rm HB},b,u}\right)^{\rm H} \left(\mathbf{F}_{b,u} - \mathbf{F}_{{\rm HB},b,u}\right) \right)\nonumber \\
    =&\mathrm{Tr}\left(  \mathbf{F}_{b,u}^{\rm H}\mathbf{F}_{b,u} - \mathbf{F}_{b,u}^{\rm H}\mathbf{F}_{{\rm HB},b,u} - \mathbf{F}_{{\rm HB},b,u}^{\rm H}\mathbf{F}_{b,u} \right. \nonumber\\
    &\left. + \mathbf{F}_{{\rm HB},b,u}^{\rm H}\mathbf{F}_{{\rm HB},b,u} \right).
    \end{align}
    Substituting (\ref{rewrite_1term})-(\ref{rewrite_4term}) into (\ref{sub_probl_F_Lagrangian_v2}) and neglecting the irrelevant terms, we can prove Proposition 1.
    \section{Proof of Lemma 1}

    Recalling the definition of $\mathbf{Q}_{b,u} = \omega_u\widehat{\mathbf{A}}_{b,u} \widehat{\mathbf{A}}_{b,u}^{\rm H}$ in Proposition 1, the rank of $\mathbf{Q}_{b,u}$ depends on $\widehat{\mathbf{A}}_{b,u}$, which is the submatrix of $\mathbf{A}_{u}^{\frac{1}{2}}$.
    Therefore, we first investigate the rank of $\mathbf{A}_{u}$.
    We decompose $\mathbf{W}_u$ as $\mathbf{W}_u = \mathbf{W}_u^{\frac{1}{2}} \left( \mathbf{W}_u^{\frac{1}{2}} \right)^{\rm H}$ and define the SVD of $\mathbf{H}_{u}^{\text{H}}\mathbf{G}_u\mathbf{W}_u^{\frac{1}{2}}$ as $\mathbf{H}_{u}^{\text{H}}\mathbf{G}_u\mathbf{W}_u^{\frac{1}{2}} = \widehat{\mathbf{U}}_{u}\widehat{\mathbf{\Sigma}}_{u}\widehat{\mathbf{V}}_{u}^{\rm H} = \widehat{\mathbf{U}}_{u}\begin{bmatrix} \overline{\mathbf{\Sigma}}_{u} \\ \mathbf{0}_{\left(BN_t-N_s\right)\times N_s} \end{bmatrix}\widehat{\mathbf{V}}_{u}^{\rm H}$, where $\overline{\mathbf{\Sigma}}_{u} = \text{diag}\left(\eta_{u,1},\eta_{u,2},\cdots,\eta_{u,N_s} \right) \in \mathbb{C}^{N_s \times N_s}$ is composed of all non-zero singular values of $\mathbf{H}_{u}^{\text{H}}\mathbf{G}_u\mathbf{W}_u^{\frac{1}{2}}$ since $\mathbf{W}_u^{\frac{1}{2}}$ is of size $N_s \times N_s$.
    Then we have
    \begin{align}\label{rank_eq_v1}
    \mathbf{A}_u = & \mathbf{H}_u^{\rm H}\mathbf{G}_u\mathbf{W}_u\mathbf{G}_u^{\rm H}\mathbf{H}_u \nonumber \\
    = & \mathbf{H}_u^{\rm H}\mathbf{G}_u\mathbf{W}_u^{\frac{1}{2}} \left(\mathbf{W}_u^{\frac{1}{2}}\right)^{\text{H}} \mathbf{G}_u^{\rm H}\mathbf{H}_u \nonumber \\
    = & \widehat{\mathbf{U}}_{u}\widehat{\mathbf{\Sigma}}_{u} \widehat{\mathbf{\Sigma}}_{u}^{\text{T}}\widehat{\mathbf{U}}_{u}^{\rm H} \nonumber \\
    = & \widehat{\mathbf{U}}_{u} \begin{bmatrix} \overline{\mathbf{\Sigma}}_{u} \overline{\mathbf{\Sigma}}_{u}^{\text{T}} & \mathbf{0}_{ N_s \times \left(BN_t-N_s\right)} \\ \mathbf{0}_{\left(BN_t-N_s\right)\times N_s} & \mathbf{0}_{ \left(BN_t-N_s\right) \times \left(BN_t-N_s\right)} \end{bmatrix} \widehat{\mathbf{U}}_{u}^{\rm H},
    \end{align}
    whose rank is $N_s$.
    Afterwards, we have
    \begin{align}\label{rank_eq_v2}
    \mathbf{A}_u^{\frac{1}{2}} = \widehat{\mathbf{U}}_u \begin{bmatrix} \ddot{\mathbf{\Sigma}}_{u} & \mathbf{0}_{ N_s \times \left(BN_t-N_s\right)} \\ \mathbf{0}_{\left(BN_t-N_s\right)\times N_s} & \mathbf{0}_{ \left(BN_t-N_s\right) \times \left(BN_t-N_s\right)} \end{bmatrix} \widehat{\mathbf{U}}_u^{\text{H}},
	\end{align}
    where $\ddot{\mathbf{\Sigma}}_{u} = \text{diag}\left(\sqrt{ \eta_{u,1}^2 },\sqrt{ \eta_{u,2}^2 },\cdots,\sqrt{ \eta_{u,N_s}^2 } \right) \in \mathbb{C}^{N_s \times N_s}$.
    Furthermore, we re-express (\ref{rank_eq_v2}) as
    \begin{align}\label{rank_eq_v3}
    & \mathbf{A}_u^{\frac{1}{2}} = \nonumber \\
    & \widehat{\mathbf{U}}_u \begin{bmatrix} \begin{bmatrix} \ddot{\mathbf{\Sigma}}_{u} & \mathbf{0}_{ N_s \times \left(N_t-N_s\right)} \\ \mathbf{0}_{\left(N_t-N_s\right)\times N_s} & \mathbf{0}_{ \left(N_t-N_s\right) \times \left(N_t-N_s\right)} \end{bmatrix} & \mathbf{0}_{ N_t \times \left(B-1\right)N_t} \\ \mathbf{0}_{\left(B-1\right)N_t \times N_t} & \mathbf{0}_{ \left(B-1\right)N_t \times \left(B-1\right)N_t} \end{bmatrix}  \nonumber\\
    & \cdot \widehat{\mathbf{U}}_u^{\text{H}},
	\end{align}
    and represent $\widehat{\mathbf{U}}_u$ as
    \begin{align}\label{rank_eq_v4}
    \widehat{\mathbf{U}}_u = \begin{bmatrix} \widehat{\mathbf{U}}_u^{\left(1,1\right) } & \widehat{\mathbf{U}}_u^{\left(1,2\right) } & \cdots & \widehat{\mathbf{U}}_u^{\left(1,B\right) } \\
    \vdots & \vdots & \ddots & \vdots \\
    \widehat{\mathbf{U}}_u^{\left(B,1\right) } & \widehat{\mathbf{U}}_u^{\left(B,2\right) } & \cdots & \widehat{\mathbf{U}}_u^{\left(B,B\right) } \end{bmatrix},
	\end{align}
    where $\widehat{\mathbf{U}}_u^{\left(i,j\right) } \in \mathbb{C}^{N_t \times N_t}$.
    Upon combining (\ref{rank_eq_v3}) and (\ref{rank_eq_v4}), we have
    \begin{align}\label{rank_eq_v5}
    \mathbf{A}_u^{\frac{1}{2}} = \begin{bmatrix} \mathbf{A}_u^{\frac{1}{2}\left(1,1\right) } & \mathbf{A}_u^{\frac{1}{2}\left(1,2\right) } & \cdots & \mathbf{A}_u^{\frac{1}{2}\left(1,B\right)} \\
    \vdots & \vdots & \ddots & \vdots \\
    \mathbf{A}_u^{\frac{1}{2}\left(B,1\right)} & \mathbf{A}_u^{\frac{1}{2}\left(B,2\right)} & \cdots & \mathbf{A}_u^{\frac{1}{2}\left(B,B\right) } \end{bmatrix},
	\end{align}
    where $\mathbf{A}_u^{\frac{1}{2}\left(i,j\right)} = \widehat{\mathbf{U}}_u^{\left(i,1\right) } \begin{bmatrix} \ddot{\mathbf{\Sigma}}_{u} & \mathbf{0}_{ N_s \times \left(N_t-N_s\right)} \\ \mathbf{0}_{\left(N_t-N_s\right)\times N_s} & \mathbf{0}_{ \left(N_t-N_s\right) \times \left(N_t-N_s\right)} \end{bmatrix} $ $\left(\widehat{\mathbf{U}}_u^{\left(j,1\right) }\right)^{\text{H}} $.
    Recalling the definition of $\widehat{\mathbf{A}}_{b,u}$, we can acquire the expression of $\widehat{\mathbf{A}}_{b,u}$ from (\ref{rank_eq_v5}) as
    \begin{align}\label{rank_eq_v6}
    \widehat{\mathbf{A}}_{b,u} = & \begin{bmatrix} \mathbf{A}_u^{\frac{1}{2}\left(b,1\right)} & \mathbf{A}_u^{\frac{1}{2}\left(b,2\right)} \cdots \mathbf{A}_u^{\frac{1}{2}\left(b,B\right)} \end{bmatrix} \nonumber\\
    = &\widehat{\mathbf{U}}_u^{\left(b,1\right) } \begin{bmatrix} \ddot{\mathbf{\Sigma}}_{u} & \mathbf{0}_{ N_s \times \left(N_t-N_s\right)} \\ \mathbf{0}_{\left(N_t-N_s\right)\times N_s} & \mathbf{0}_{ \left(N_t-N_s\right) \times \left(N_t-N_s\right)} \end{bmatrix} \nonumber\\
    & \begin{bmatrix} \left(\widehat{\mathbf{U}}_u^{\left(1,1\right) }\right)^{\text{H}} & \left(\widehat{\mathbf{U}}_u^{\left(2,1\right) }\right)^{\text{H}} & \cdots & \left(\widehat{\mathbf{U}}_u^{\left(B,1\right) }\right)^{\text{H}} \end{bmatrix} \nonumber \\
    \overset{\Delta}{=} & \widetilde{\mathbf{A}}_1 \widetilde{\mathbf{A}}_2 \widetilde{\mathbf{A}}_3,
	\end{align}
    where we define $\widetilde{\mathbf{A}}_1 = \widehat{\mathbf{U}}_u^{\left(b,1\right) }$, $\widetilde{\mathbf{A}}_2 = \begin{bmatrix} \ddot{\mathbf{\Sigma}}_{u} & \mathbf{0}_{ N_s \times \left(N_t-N_s\right)} \\ \mathbf{0}_{\left(N_t-N_s\right)\times N_s} & \mathbf{0}_{ \left(N_t-N_s\right) \times \left(N_t-N_s\right)} \end{bmatrix}$ and \\ $\widetilde{\mathbf{A}}_3 =  \begin{bmatrix} \left(\widehat{\mathbf{U}}_u^{\left(1,1\right) }\right)^{\text{H}} & \left(\widehat{\mathbf{U}}_u^{\left(2,1\right) }\right)^{\text{H}} & \cdots & \left(\widehat{\mathbf{U}}_u^{\left(B,1\right) }\right)^{\text{H}} \end{bmatrix}$.
    Adopting the Sylvester inequality, i.e., $\text{rank}\left( \mathbf{X} \right) + \text{rank}\left( \mathbf{Y} \right) - x_{\rm col} \leq \text{rank}\left( \mathbf{X}\mathbf{Y} \right) \leq \min\left\{\text{rank}\left( \mathbf{X} \right) , \text{rank}\left( \mathbf{Y} \right)\right\}$
    with $x_{\rm col}$ denoting the number of columns in $\mathbf{X}$ \cite{MatrixAnalysis}, we have $\text{rank}\left( \widehat{\mathbf{A}}_{b,u} \right) \leq \text{rank} \left(  \widetilde{\mathbf{A}}_2 \right) = N_s$.
    Moreover, we have
    \begin{align}\label{rank_eq_v6pp1}
    \text{rank}\left( \widehat{\mathbf{A}}_{b,u} \right) \geq & \text{rank}\left( \widetilde{\mathbf{A}}_1\right) + \text{rank}\left(  \widetilde{\mathbf{A}}_2 \widetilde{\mathbf{A}}_3 \right) - N_t \nonumber \\
    = & N_t + \text{rank}\left(  \widetilde{\mathbf{A}}_2 \widetilde{\mathbf{A}}_3 \right) - N_t \nonumber\\
    \geq & \text{rank}\left(  \widetilde{\mathbf{A}}_2 \right) + \text{rank}\left( \widetilde{\mathbf{A}}_3 \right) - N_t \nonumber\\
    = & N_s + N_t - N_t \nonumber \\
    = & N_s.
	\end{align}
    Then we have $\text{rank}\left(  \widehat{\mathbf{A}}_{b,u} \right) = N_s$.
    Finally, by adopting the property that $ \text{rank}\left( \mathbf{X}\mathbf{X}^{\rm H} \right) = \text{rank}\left( \mathbf{X} \right) $ \cite{MatrixAnalysis}, we have
    \begin{align}\label{rank_eq_v7}
    \text{rank}\left(\mathbf{Q}_{b,u} \right) = \text{rank}\left(\widehat{\mathbf{A}}_{b,u} \widehat{\mathbf{A}}_{b,u}^{\rm H}\right)  =  \text{rank}\left(\widehat{\mathbf{A}}_{b,u}\right) = N_s.
	\end{align}
    Hence we can prove the Lemma.
    \end{appendices}
	\bibliographystyle{IEEEtran}
	\bibliography{IEEEabrv,Refference}

\begin{IEEEbiography}[{\includegraphics[width=1in,height=1.25in,clip,keepaspectratio]{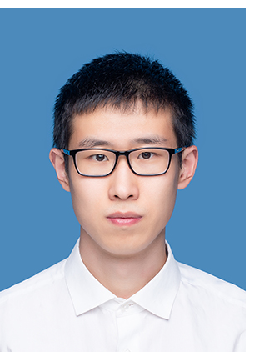}}]
		{Chenghao Feng} received the B.E. degree from the Beijing Institute of Technology, Beijing, China, in 2017, where he is currently pursuing the Ph.D. degree with the School of Information and Electronics. His current research interests include massive MIMO, mmWave/THz communications, energy-efficient communications, intelligent reflecting surface and networks.
\end{IEEEbiography}

\begin{IEEEbiography}[{\includegraphics[width=1in,height=1.25in,clip,keepaspectratio]{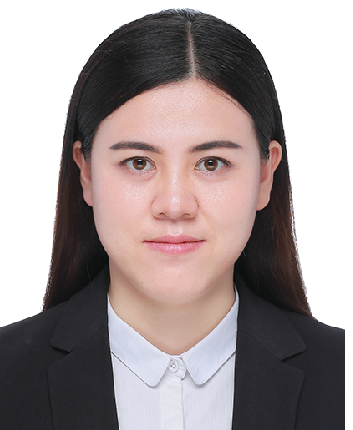}}]
		{Wenqian Shen} received the B.S. degree from Xi'an Jiaotong University, Shaanxi, China in 2013 and the Ph.D. degree from Tsinghua University, Beijing, China in 2018. She is currently an associate professor with the School of Information and Electronics, Beijing Institute of Technology, Beijing, China. Her research interests include massive MIMO and mmWave/THz communications. She has published several journal and conference papers in IEEE Transaction on Signal Processing, IEEE Transaction on Communications, IEEE Transaction on Vehicular Technology, IEEE ICC, etc. She has won the IEEE Best Paper Award at the IEEE ICC 2017.
\end{IEEEbiography}

	\begin{IEEEbiography}[{\includegraphics[width=1in,height=1.25in,clip,keepaspectratio]{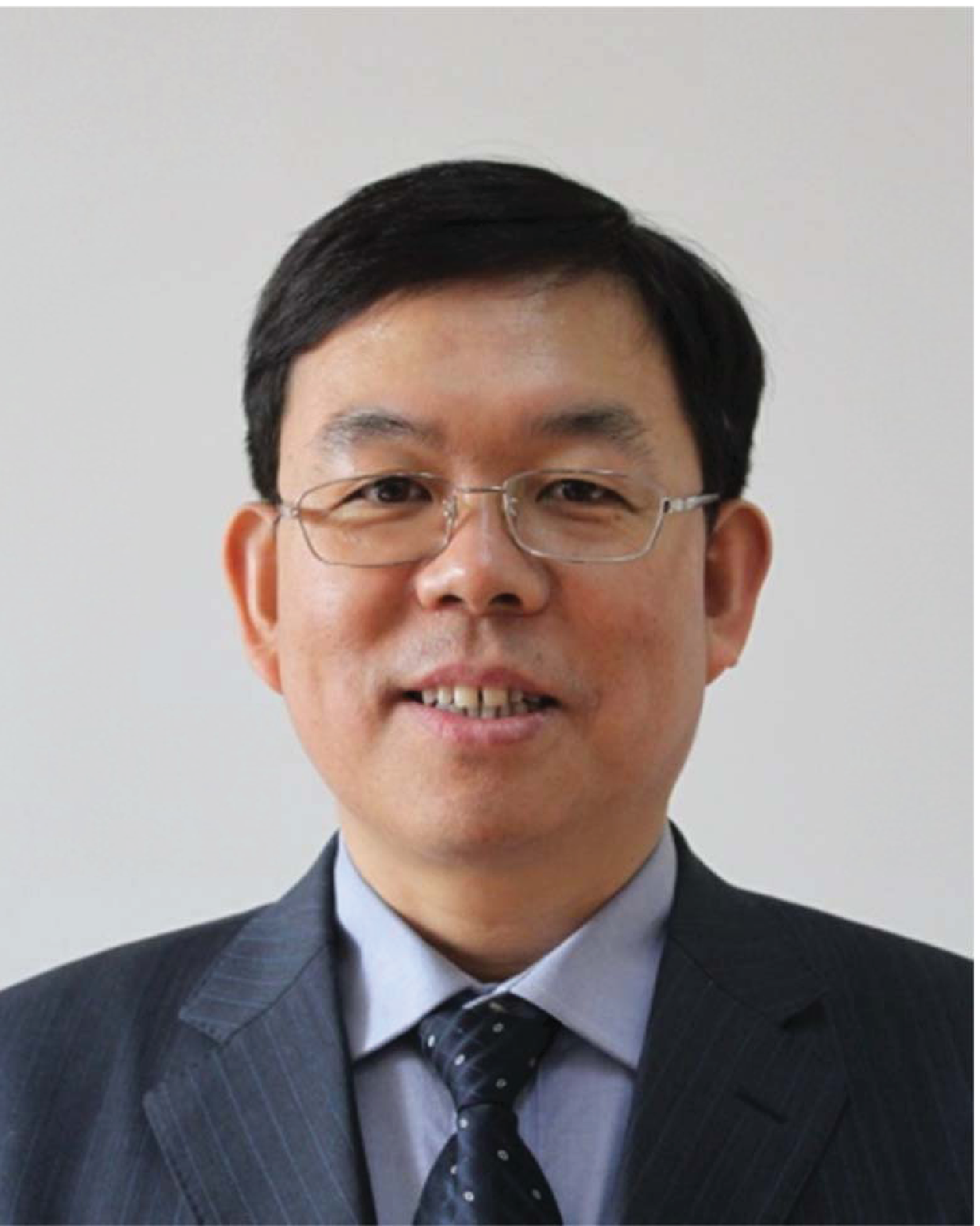}}]
		{Jianping An} (M'08) received the B.E. degree from Information Engineering University in 1987, and the M.S. and Ph.D. degrees from Beijing Institute of Technology, in 1992 and 1996, respectively. Since 1996, he has been with the School of Information and Electronics, Beijing Institute of Technology, where he now holds the post of Full Professor. From 2010 to 2011, he was a Visiting Professor at University of California, San Diego. He has published more than 150 journal and conference articles and holds (or co-holds) more than 50 patents. He has received various awards for his academic achievements and the resultant industrial influences, including the National Award for Scientific and Technological Progress of China (1997) and the Excellent Young Teacher Award by the China's Ministry of Education (2000). Since 2010, he has been serving as a Chief Reviewing Expert for the Information Technology Division, National Scientific Foundation of China. Prof. An's current research interest is focused on digital signal processing theory and algorithms for communication systems.
	\end{IEEEbiography}

\begin{IEEEbiography}[{\includegraphics[width=1in,height=1.25in,clip,keepaspectratio]{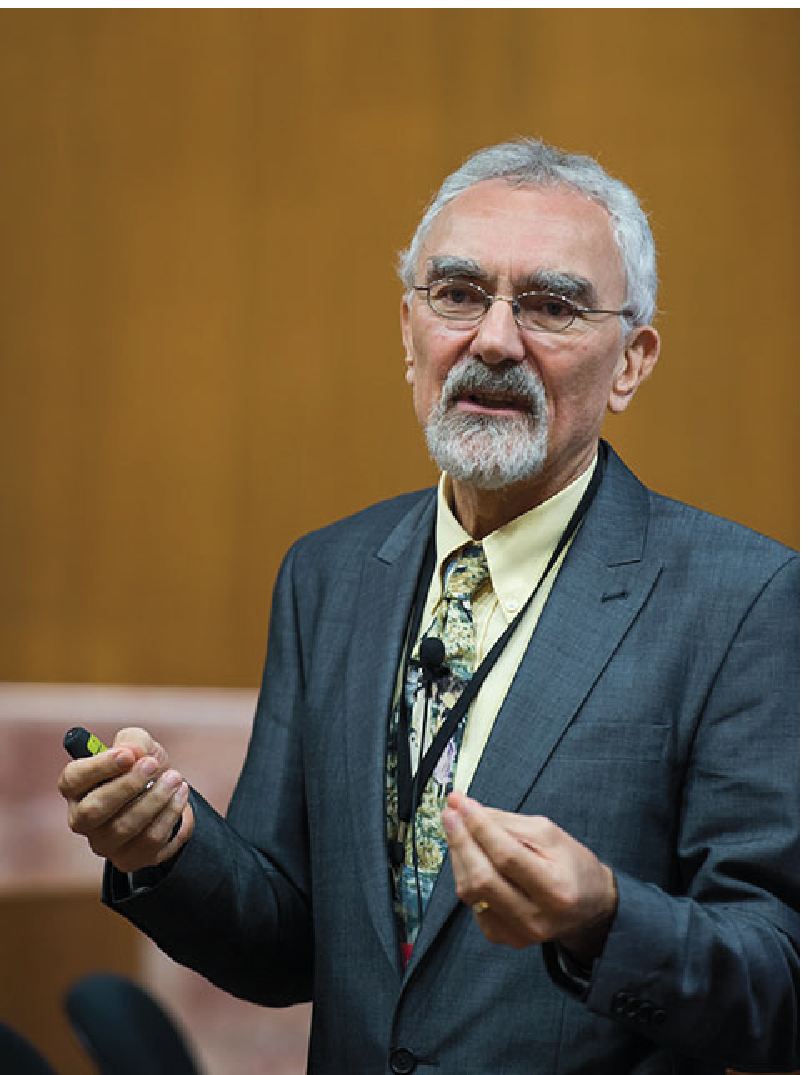}}]
		{Lajos Hanzo}  (http://www-mobile.ecs.soton.ac.uk,
https://en.wikipedia.org/wiki/Lajos\_Hanzo) (FIEEE'04) received his
Master degree and Doctorate in 1976 and 1983, respectively from the
Technical University (TU) of Budapest. He was also awarded the Doctor of
Sciences (DSc) degree by the University of Southampton (2004) and
Honorary Doctorates by the TU of Budapest (2009) and by the University
of Edinburgh (2015).  He is a Foreign Member of the Hungarian Academy of
Sciences and a former Editor-in-Chief of the IEEE Press.  He has served
several terms as Governor of both IEEE ComSoc and of VTS.  He has
published 1970 contributions at IEEE Xplore, 19 Wiley-IEEE Press books
and has helped the fast-track career of 123 PhD students. Over 40 of
them are Professors at various stages of their careers in academia and
many of them are leading scientists in the wireless industry. He is also
a Fellow of the Royal Academy of Engineering (FREng), of the IET and of
EURASIP.
\end{IEEEbiography}
\end{document}